\documentclass[11pt]{article}
\usepackage[UKenglish]{babel}
\usepackage[a4paper, margin=1in]{geometry}

\usepackage{hyperref}
\usepackage{graphicx}
\usepackage{caption}
\usepackage{subcaption}
\usepackage{amsmath}
\usepackage{amssymb}
\usepackage{mathtools}
\usepackage{siunitx}
\usepackage{multirow}
\usepackage{xcolor}
\usepackage{ctable}
\usepackage{bm}
\usepackage{upgreek}
\usepackage{xcolor}

\graphicspath{{figures/}}

\title{3D variational autoencoder for fingerprinting microstructure volume elements}
\author{Michael D. White$^{1}$\footnote{
    Corresponding author \newline \textit{Email address:} mike.white@ukaea.uk (M. White)},
    Michael D. Atkinson$^{1}$,
    Adam J. Plowman$^{1}$,
    Pratheek Shanthraj$^{1}$ \\ \\
    \small $^{1}$\textit{UKAEA, Culham Campus, Abingdon, Oxfordshire, OX14 3DB} \\
    }
\date{\small \today}

\begin{document}

\maketitle

\noindent\rule{\textwidth}{.5pt}
\begin{abstract}
Microstructure quantification is an important step towards establishing structure-property relationships in materials.
Machine learning-based image processing methods have been shown to outperform conventional image processing techniques and are increasingly applied to microstructure quantification tasks.
In this work, we present a 3D variational autoencoder (VAE) for encoding microstructure volume elements (VEs) comprising voxelated crystallographic orientation data.
Crystal symmetries in the orientation space are accounted for by mapping to the crystallographic fundamental zone as a preprocessing step, which allows for a continuous loss function to be used and improves the training convergence rate.
The VAE is then used to encode a training set of VEs with an equiaxed polycrystalline microstructure with random texture.
Accurate reconstructions are achieved with a relative average misorientation error of $3 \times 10^{-2}$, on the test dataset, for a continuous latent space with dimension 256.
We show that the model generalises well to microstructures with textures, grain sizes and aspect ratios outside the training distribution.

The main aim of this paper is to present the 3D VAE as an application-agnostic method for parameterising microstructure for input into downstream tasks where microstructure dependence is required.
As a proof of concept, a simple surrogate model for uniaxial crystal plasticity (CP) simulations, with a fixed load path and microstructural dependence is presented.
Microstructural fingerprints, obtained by encoding VEs with the trained VAE encoder, parameterise the VEs in a low-dimensional latent space and are stored alongside the volume-averaged stress response, at each strain increment.
This is then used to train a fully connected neural network mapping the input fingerprint to the resulting stress response, which acts as a surrogate model for the CP simulation.
The fingerprint-based surrogate model is shown to accurately predict the microstructural dependence in the CP stress response, with a mean relative error (MSE) of 2.75 MPa on unseen test data.
This approach offers a significant speed-up on the order of $10^{8}$ for a stress-strain curve prediction, compared to running a CP simulation.

\end{abstract}
\noindent\rule{\textwidth}{.5pt}

\newpage


\section{Introduction}
\label{sec:intro}

To understand the macroscopic behaviour of materials beyond the regimes that have been probed experimentally, it is important to consider structure-property relations from the atomistic scale up through the microstructural and continuum length scales~\cite{van_der_giessen_roadmap_2020, mianroodi_modeling_2022, diehl_solving_2020}.
Addressing this key challenge is hindered by the severe computational costs associated with conventional concurrent~\cite{lu_electrons_2006} and coarse-grained multiscale approaches~\cite{geers_multi-scale_2010}.
More recently, deep learning-based methods have shown great promise in emulating coarse-grained reduced order models whilst offering significant computational speed-up~\cite{alber_integrating_2019,bishara_state---art_2023}.
At the microstructural scale, deep learning-based reduced order models for crystal plasticity (CP) have been proposed, which utilise representative volume elements for a given material, and predict homogenized stress responses for given load paths~\cite{bonatti_cp-fft_2022, liu_learning_2023, ghavamian_accelerating_2019, wu_recurrent_2020}.
While these approaches have been shown to provide accurate predictions, training is often limited to smaller volume elements (VEs), that are not representative of the microstructure distribution~\cite{fernandez-zelaia_self-supervised_2024}.
As a result, such model responses can have a much higher variance and become heavily microstructure dependent.

One of the key challenges with including microstructure dependence is finding an efficient method for parameterising microstructure~\cite{Burnett2019, Agrawal2016, McDowell2016, bostanabad_computational_2018}.
Metrics can be applied, such as mean grain size and variance, with aspect ratios and phase fractions, but these metrics are limited and do not generalise well to arbitrary microstructures~\cite{hahn_grain-size_2015}.
Such metrics also lack the uniqueness for a given realisation of a VE, but rather capture the statistics that the VE was sampled from.
Similarly, $n$-point statistics can provide more detailed statistical descriptors, including spatial correlations, of a wider range of microstructures~\cite{torquato_statistical_2002, kalidindi_microstructure_2011, niezgoda_optimized_2010, fullwood_microstructure_2008}, and have been successfully used to describe structure-property relationships~\cite{hu_microstructure-informatic_2021} and microstructure evolution~\cite{chen_hierarchical_2019}.
In special cases, such as where single-phase polycrystalline VEs are generated through tessellations, it is possible to get a unique descriptor, or fingerprint, for a given VE by concatenating the seed attributes with the crystallographic orientation of each grain used to generate the VE.
In more general cases, such as multiphase materials and complex grain structures, the VE generation may not be dependent on input parameters by a simple function or, possibly, even be unknown.
Therefore, a more generalisable method is required for fingerprint construction.

To address this task, a variety of methods have been proposed over the last few decades.
Local descriptors, centred at key points in an image, can be aggregated with statistical methods to construct distributions of typical features that can quantify a microstructure~\cite{DeCost2015, Decost2017, White2023} and this has been shown to provide a useful parameterisation for classification tasks.
Linear projection, through principal component analysis (PCA), has also been used to develop physically meaningful descriptors~\cite{suh_application_2002}.
More recently, attention has shifted towards convolutional neural networks (CNNs)~\cite{Fukushima1980}.
Pretrained networks for popular architectures, such as AlexNet~\cite{Krizhevsky2012}, are readily available and have been shown to adapt well to microstructure, despite having been trained on natural images~\cite{Holm2020}.
Fine-tuning can then be applied to further optimise a pretrained network for a specific task, such as constructing structure-property relationships~\cite{Li2018}.
One of the drawbacks with a standard CNN is the requirement for a predefined task and labelled data to train the network.
On the other hand, an unsupervised method for training a CNN is the variational autoencoder (VAE)~\cite{Kingma2013}.
VAEs are composed of a pair of CNNs, one for encoding the input and another that aims to reconstruct the input from the encoded representation.
Once trained, the encoder portion of the VAE has been shown to provide suitable microstructure parameterisation for structure-property predictions~\cite{Cang2018, Kim2021}.
The advantage here is that the fingerprinting process is abstracted away from the predictive model, providing an application agnostic parameterisation for microstructure that can be fed into arbitrary downstream tasks, which is the main focus of this paper.

Here, we consider a 3D VAE and treat the trained encoder as a map from VE to fingerprint.
The encoder is based on the popular ResNet~\cite{he_deep_2015} architecture and utilises 3D convolutions at each layer.
VAEs have been shown to be a suitable method for constructing meaningful descriptors of microstructure and common microstructural features, such as phase fractions and grain morphology, can be correlated with position in the latent space~\cite{White2024, oommen_learning_2022, desai_trade-offs_2024}.
The VAE is trained on randomly generated VEs with orientations randomly sampled across the Euler angle domain to capture it in its entirety during training, with a crystallographic fundamental zone used to remove symmetries~\cite{krakow_three-dimensional_2017}.
Single phase microstructures are considered with equiaxed grains, although the methods presented can be extended to multiphase VEs, as symmetries are accounted for during preprocessing and not during training.

The trained VAE is then used to generate a dataset of microstructural fingerprints, and DAMASK~\cite{roters_damask_2019} is used to perform CP simulations of uniaxial tension applied to the corresponding VEs.
The resulting fingerprints and volume-averaged stress responses at each strain increment are then used to train and validate a fully connected neural network surrogate for the CP model.
There are a wide range of options for the surrogate model, but a fully connected architecture was selected as there are no limitations on suitable input size (unlike Gaussian process regression).
The time-dependence of the loading is implicit in the model, as this is constant for each simulation, so a fixed number of output nodes could be used.
It is important to note here that this is a simple model and is presented more as a proof of concept to illustrate the effectiveness of VAE encoding for parameterising microstructure for a downstream task which possesses microstructural dependence.
To extend this work to operate on an arbitrary load path, recurrent neural networks (RNNs) and/or transformers would need to be considered.
Therefore, in this case, the full pipeline consists of VE generation, VAE fingerprinting, CP simulations of uniaxial tension and, finally, a fully connected network to predict stress response.
However, more generally, the surrogate model could be swapped out for a different predictive model that deals with some other material properties of interest, but the same trained VAE could be used without any additional training of that portion of the pipeline.
The latent space constructed by the trained VAE is explored with dimensionality reduction via PCA for a visual aid to observe microstructural feature clustering throughout the latent space, and spectral clustering is used to further quantify the qualitative results from the PCA.
Finally, we show that fingerprinting microstructure VEs in this way can provide a suitable descriptor for parameterising microstructure as inputs for a microstructure-dependent CP surrogate model.


\section{Materials and Methods}
\label{sec:methods}


\subsection{Datasets}
\label{subsec:datasets}

Several datasets of 3D single phase VEs were constructed using Dream3D~\cite{groeber_dream3d_2014}.
Each VE is converted into a PyTorch~\cite{Paszke2019} tensor with shape $(3, 64, 64, 64)$, i.e.\ a $64^3$ VE with 3 channels at each voxel to describe crystal orientation as an Euler angles in the Bunge convention.
The Euler angles are normalised to a range of $[0, 1]$ for training purposes, by dividing by $(2\pi, \pi, 2\pi)$ across the three channels respectively.
We also use the DAMASK Python library~\cite{otto_2025} to map the Euler angles into the crystallographic fundamental zone to remove crystal symmetries as a preprocessing step.
This requires prior knowledge about the crystal structure.
We consider cubic and hexagonal symmetries here and, as such, the trained models do not extend beyond that.
Most of the results presented herein use a cubic crystal structure.
The hexagonal close-packed (HCP) structure is introduced for the surrogate model, as it provides a more anisotropic stress response to train against.
There are several options for taking symmetry into account, but it was deemed that addressing crystallographic symmetries during preprocessing provided the most stability during training.
Alternatively, symmetries could be considered in the loss function during training.
However, we found that this caused numerical instabilities, due to enforcing a discontinuous loss function, and it can be expensive to compute at each epoch, due to the requirement for computing all symmetries at each epoch.
In theory, all symmetric variants of the training data could be computed prior to training, but it is not practical to store them in memory due to the size of the dataset.

\subsubsection{VAE Training Dataset}
\label{subsubsec:train_data}

Single phase equiaxed microstructures were generated from a distribution of the equivalent sphere diameter (ESD) to control the morphology.
The distribution parameters were selected to produce approximately 10 grains per face of the VE and are provided in Table~\ref{tab:morphology_stats}.
At this VE size, each grain has roughly 2-3 grain neighbours, which was chosen as microstructure perturbations beyond this neighbourhood had insignificant effects on the strain field within the grain.

\begin{table}[!ht]
    \centering
    \begin{tabular}{ |c|c| }
        \hline
        \textbf{Feature}  & \textbf{Value} \\
        \hline
        Crystal structure & Cubic          \\
        \hline
        Mean feature ESD  & 5.78           \\
        \hline
        Sigma             & 0.1            \\
        \hline
        Bin step size     & 0.5            \\
        \hline
        Min cut off       & 5              \\
        \hline
        Max cut off       & 5              \\
        \hline
    \end{tabular}
    \caption{Grain morphology statistics used to generate the VEs in Dream3D with $64^3$ voxels and a resolution of 0.25 units per voxel, where the units are relative to resolution per voxel.}
    \label{tab:morphology_stats}
\end{table}

Grain orientations were obtained through randomly sampling the space of all valid Euler angles, with the aim of capturing an orientation distribution without any bias towards specific textures.
These were then reduced to the crystallographic fundamental zone to remove symmetries and mitigate the risk of any many-to-one mappings that may occur in the surrogate models.
It is important to note that each VE individually is not representative of the microstructure distribution being sampled from, but rather a small subset.
This enforces more variability across the dataset with the objective of microstructural dependence in mind, however, this leads to the requirement for large datasets for training purposes.
In this case, we generate a dataset of 100,000 VEs for training and 10,000 for model validation.

\subsubsection{Edge Case Datasets}
\label{subsubsec:test_data}

To test how well the trained VAE can adapt and generalise to unseen data, various datasets of edge cases were created, targeting morphological features and specific texture components.
For the morphology edge cases, new distributions of grain sizes were created to provide VEs with either large or small grains, relative to the training data.
Parameters to generate the distributions remain consistent with those provided in Table~\ref{tab:morphology_stats}, except the ESD means, which were set to 12.24 and 2.73, respectively.
We also considered the case of elongated grains, in which the aspect ratio was set to 2:1:1, with respect to the $(x, y, z)$ directions, and the ESD mean was set to 9.54.
This is slightly larger than the training data ESD mean, but this adjustment compensates for the elongation reducing the diameter at the cross-section of each grain.

For the edge cases in orientation space, four common textures were selected.
Namely, brass, copper, cube and Goss.
The mean Euler angle for each of these is shown in Table~\ref{tab:texture_stats}.
In each case, a random morphology was generated (following the same distribution as the training data) and the texture was sampled from normal distributions centred at each Euler angle with a variance of $2^\circ$.
The sampled Euler angles were mapped into the crystallographic fundamental zone, as before.
Care was also taken to avoid any periodicity effects in the cases where the mean was zero, resulting in any added noise being restricted to be positive in these cases, as this was not fully accounted for by mapping to the fundamental zone.
It should be noted that this restriction leads to a lower variance in the sampled orientations that are close to zero which, in turn, leads to tighter clustering in the latent space for VEs containing Euler angles close to zero.

\begin{table}[!ht]
    \centering
    \begin{tabular}{ |c|c| }
        \hline
        \textbf{Texture} & \textbf{Orientation} \\
        \hline
        Brass            & ($35^\circ$, $45^\circ$, $0^\circ$)          \\
        \hline
        Copper           & ($90^\circ$, $35^\circ$, $45^\circ$)         \\
        \hline
        Cube             & ($0^\circ$, $0^\circ$, $0^\circ$)            \\
        \hline
        Goss             & ($0^\circ$, $45^\circ$, $0^\circ$)           \\
        \hline
    \end{tabular}
    \caption{Euler angle triplets, in degrees, used to represent each texture included as an edge case. Each orientation is sampled with means as shown here, with a variance of $2^\circ$.}
    \label{tab:texture_stats}
\end{table}


\subsection{Latent Variable Models}
\label{subsec:latent_var_model}

Given a set of microstructure VE realisations, $x$, a latent variable model aims to construct the marginal likelihood, $p(x)$, and the posterior distribution, $p(z | x)$.
The prior distribution, $p(z)$, is user-defined and is typically set to a standard unit normal distribution.
We can also define some decoder with learnable parameters to give the likelihood $p(x| z)$.
A latent variable model can then be trained to give us an approximation to the posterior distribution, $p(x| z)$, which can then be used to encode a microstructure.

\subsubsection{Variational Autoencoders (VAEs)}
\label{subsubsec:vaes}

Variational autoencoders (VAEs) are a subset of latent variable models that are composed of a pair of CNNs.
The first CNN acts as an encoder, downsampling the input data to some hidden state, which is then used to sample a set of means, $\mu_i$, and variances, $\sigma_i$.
These are then used to construct a latent representation, $z$, of the input data, $x$, given by
\begin{equation}
    z_i = \mu_i + e^{\frac{1}{2} \ln(\sigma_i) \mathcal{N}(0, 1)},
\end{equation}
where $\mathcal{N}(0, 1)$ is our chosen prior distribution.
The random noise is sampled here to aid generalisation of the model to unseen data.
Once the model is trained, to extract a fingerprint from an input VE, it is usual to discard this noise term and treat $\mu$ as the fingerprint.
The $z$ is then supplied as an input to another CNN, which acts as a decoder.
This can be thought of as an approximation to the inverse operator of the encoder, which employs convolutional transpose operators to upsample the latent representation back into the image domain.
These act in much the same way as a convolution operation, with some kernel that is convolved with an input.
However, in the transpose case, each kernel operates on a single pixel/voxel and upsamples the input to a space that is dimensionally equivalent to the kernel.

\begin{figure}[!ht]
    \centering
    \includegraphics[width=\textwidth]{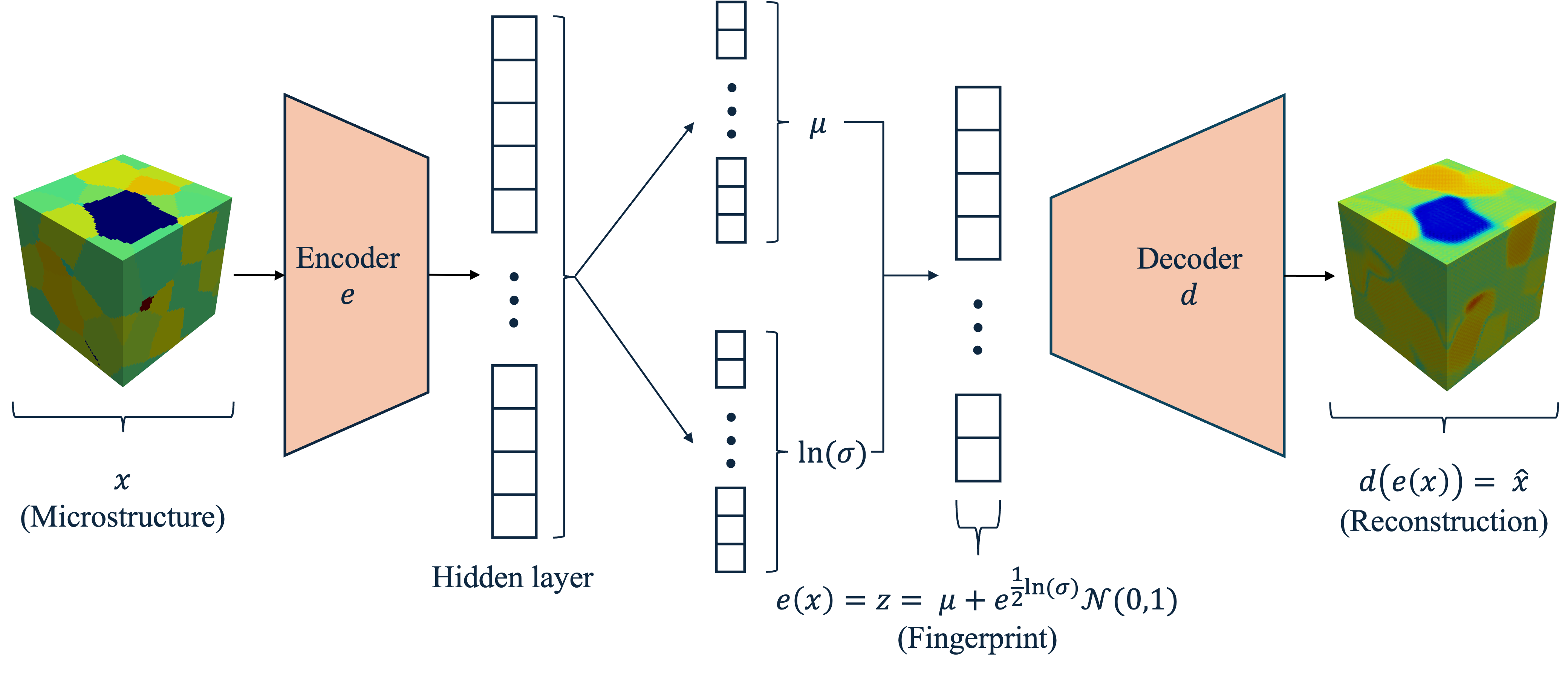}
    \caption{VAE schematic for encoding microstructure VEs using 3D convolutions.}
    \label{fig:vae_schematic}
\end{figure}

VAEs are typically applied to image data, $x \in \mathbb{R}^{m_1 \times m_2 \times C}$, where $m_1, m_2$ is the number of pixels in the image and $C$ denotes the number of channels.
To adapt a CNN for voxelated data, 3D kernels are used at each convolution layer and applied to each channel (corresponding to each normalised Euler angle) independently.
Figure~\ref{fig:vae_schematic} shows a schematic of the architecture discussed.

There are several hyperparameters that are important to tune here.
Primarily, the architectures selected for the encoder and decoder will control the feature extraction process and this will be discussed in more detail in Section~\ref{subsubsec:3dvae}.
We also need to be careful with our choice of dimensionality for the latent space, ensuring an optimal ratio between feature extraction and compression.
Finally, the size of the hidden state is also crucial, as this acts as a bottleneck in compression.
Setting this too small yields numerical instabilities that can result in random outliers appearing in the latent representations, which are easily identified through dimensionality reduction.
The effects of this can be seen in~\cite{White2024}.

\subsubsection{ResNet3D VAE}
\label{subsubsec:3dvae}

ResNet is a commonly used CNN that is composed of a series of residual blocks~\cite{he_deep_2015}.
Each block contains a convolution layer, followed by batch normalisation and ReLU activation.
This is then followed by another convolution layer and batch normalisation with the output from this being summed with the initial input into the residual block.
This serves as a so-called skip connection that allows for some layers to be effectively ignored, enabling shallow feature extraction where appropriate.
The output from this is then passed through a final ReLU activation layer.

Each layer in the ResNet is then formed by concatenating these blocks.
No downsampling happens when applying these blocks, so the user is free to choose as many or as few as they like, without needing to modify anything else about the architecture.
Between each block is a separate convolution layer for downsampling.
There are 4 layers in total and the depth of the network is controlled by specifying the number of residual blocks in each layer.
Here, we consider ResNet12, ResNet18 and ResNet34.
The number of blocks in each layer for these architectures are summarised in Table~\ref{tab:res_blocks}.

\begin{table}[!ht]
    \centering
    \begin{tabular}{ |c|c| }
        \hline
        \textbf{Architecture} & \textbf{Number of Residual Blocks} \\
        \hline
        ResNet12              & [1, 1, 1, 1]                       \\
        \hline
        ResNet18              & [2, 2, 2, 2]                       \\
        \hline
        ResNet34              & [3, 4, 6, 3]                       \\
        \hline
    \end{tabular}
    \caption{Number of blocks in each layer of the ResNet architectures tested in this study.}
    \label{tab:res_blocks}
\end{table}

During training, data is passed through the VAE to yield a fingerprint and a reconstruction.
A loss function is then applied to quantify the reconstruction accuracy, and this is utilised to inform an optimiser for backpropagation.
During backpropagation, weights are updated in the network with the aim of minimising the loss function.
Here, we use the Adam optimiser~\cite{kingma_adam_2017} with a learning rate of $1 \times 10^{-5}$ and weight decay of $1 \times 10^{-5}$.

\subsubsection{Loss Functions}
\label{subsubsec:loss_funcs}

The loss function used for a VAE is typically comprised of two terms \,---\, one for quantifying the accuracy of the reconstruction, and one to act as a regulariser on the latent space.
Typically, these are selected as the mean-squared error (MSE) loss and the Kullback-Leibler (KL) divergence, respectively~\cite{Kingma2019}.
These can work well in practice, but other options have also been proposed, such as spectral regularisation paired with binary cross-entropy (BCE) loss~\cite{Bjork2022}.
Here, the BCE loss acts as the reconstruction accuracy term, which is given as
\begin{equation}
    \mathcal{L}_{\text{BCE}}(x, \hat{x}) = - \sum_{i=1}^{m} x_i \log (\hat{x}_i) - \sum_{i=1}^{m} (1 - x_i) \log (1 - \hat{x}_i),
\end{equation}
where $x$ denotes the input VE and $\hat{x}$ denotes the corresponding reconstruction.
The spectral loss is then given by the MSE between the 3D Fourier transforms of $x$ and $\hat{x}$, i.e.\
\begin{equation}
    \mathcal{L}_{\text{FFT}}(x, \hat{x}) = \frac{1}{m} \sum_{j=1}^{m} \left[(\text{Im}\{\mathcal{F}(x)_j\} - \text{Im}\{\mathcal{F}(\hat{x})_j\})^2 + (\text{Re}\{\mathcal{F}(x)_j\} - \text{Re}\{\mathcal{F}(\hat{x})_j\})^2 \right],
\end{equation}
where $\mathcal{F}(x)$ denotes the 3D fast Fourier transform of $x$.
This acts as a regulariser, although this also operates on the reconstruction and not on the latent space directly.
As such, this term also contributes to quantifying the reconstruction accuracy.
This loss function was proposed as a method for reducing the smoothness of VAE reconstructions~\cite{Bjork2022} and seems to be particularly effective at handing microstructural image data, due to the repeating nature of microstructural data across the image domain.
The full spectral loss function, $\mathcal{L}(x, \hat{x})$, is then given as a weighted sum between the two terms, such that
\begin{equation}
    \label{eq:loss}
    \mathcal{L}(x, \hat{x}) = \alpha \left[ \mathcal{L}_{\text{BCE}}(x, \hat{x}) \right] + (1 - \alpha) \left[ \mathcal{L}_{\text{FFT}}(x, \hat{x}) \right],
\end{equation}
where $\alpha$ is a hyperparameter that can be tuned.
For the results provided in this paper, a value of $\alpha = 0.999$ is used throughout, as the Fourier term yields values that generally are on the order of $1 \times 10^{4}$ to $1 \times 10^{5}$ times larger than the BCE term.


\subsection{Crystal Plasticity (CP) Model}
\label{subsec:cp_model}

The open source CP simulation software DAMASK~\cite{roters_damask_2019} was used alongside MatFlow~\cite{plowman_novel_2023} for managing high-concurrency workflows to generate the training data for the surrogate model.
The workflow generates a VE using Dream3D, as described in Section~\ref{subsubsec:train_data}, post-processes the output into a format readable by DAMASK and performs a CP simulation with uniaxial loading along the $x$-axis.
Each VE is deformed to a strain of 0.01 at a rate of $1 \times 10^{-3}$ over 20 increments, to characterize the plastic yielding behaviour of the VE.
The VE is also passed through the trained VAE to construct a microstructural fingerprint in parallel with performing the CP simulation.
Finally, the workflow output is processed to yield a dataset containing each VE's fingerprint, and the associated volume-averaged stress at each strain increment.
The workflow was executed to run 50,000 simulations, 48,401 of which converged to a solution, with some runs not converging due to the severe plastic anisotropy of HCP magnesium.
This was split into a training set of 40,000, with the rest reserved for model validation.
Each simulation was run across 4 cores on Intel Sapphire Rapids CPUs, with a wall-clock time of $\sim 10$ minutes per simulation.
The volume-averaged stress responses were normalised using min-max scaling, across the whole dataset, to a range of [0, 1], prior to splitting the dataset into training and validation partitions.
Single crystal parameters for a magnesium phenomenological CP model~\cite{liu_integrated_2018} were used to generate this dataset, as it provides a relatively large spread in stress response compared to copper due to higher anisotropy.
The anisotropy comes from magnesium having a HCP crystal structure, and so the VAE was retrained, in the same manner as previously described and with the relevant fundamental zone, before being used for fingerprint construction.


\subsection{Surrogate Model}
\label{subsec:surroagate}

\begin{figure}[!ht]
    \centering
    \includegraphics[width=\textwidth]{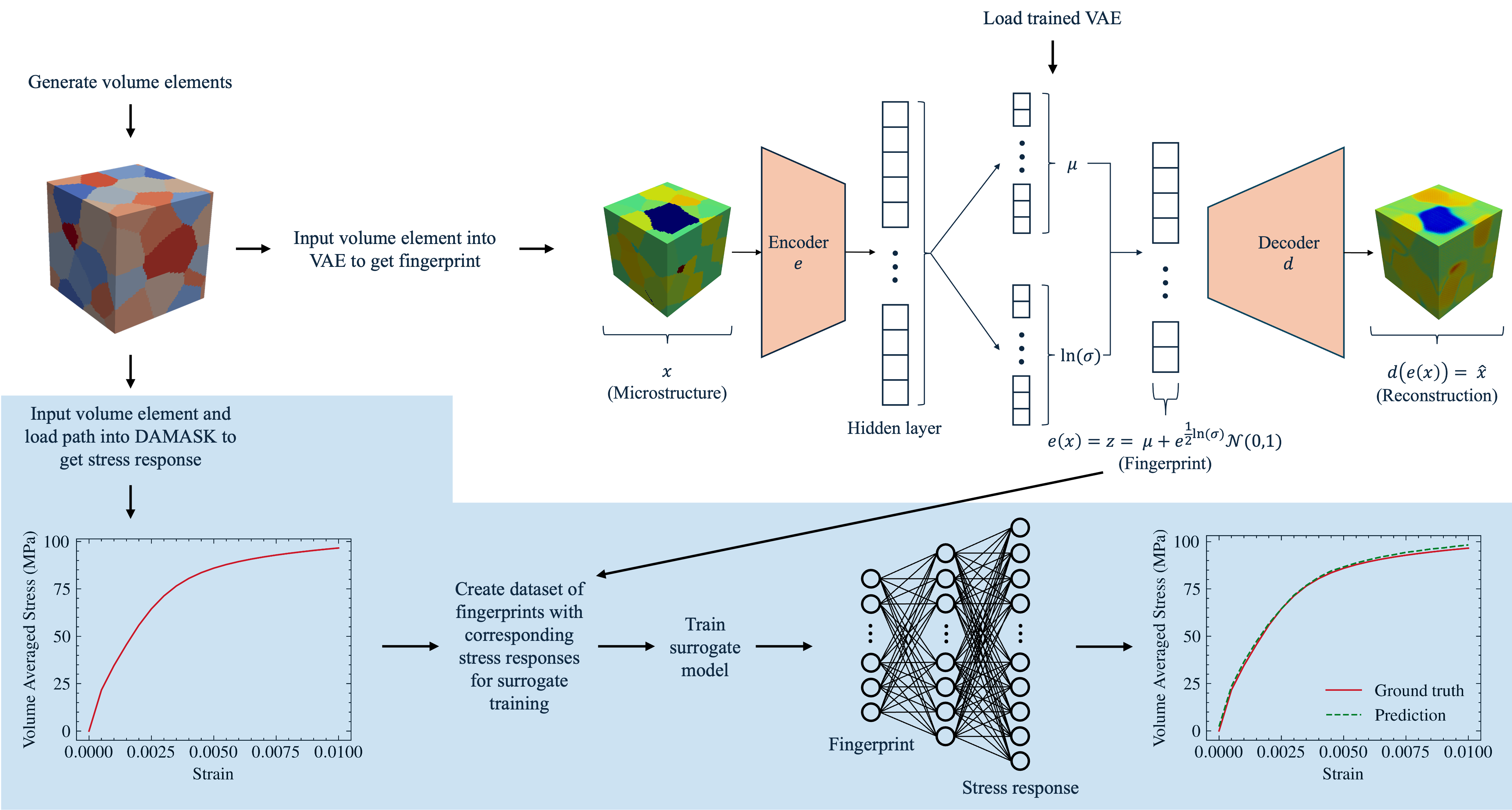}
    \caption{Schematic of the full pipeline of the VAE fingerprinting process with an example surrogate model. The region highlighted in blue is interchangeable with other microstructure-dependent surrogate model options, whilst retaining the trained VAE and fingerprinting processes.}
    \label{fig:pipeline_schematic}
\end{figure}

As discussed, the VAE fingerprinting process is intentionally distinct from the surrogate model, with the aim being to construct an application agnostic parameterisation for microstructure, rather than training a fully end-to-end model from microstructure to stress response.
Here, we consider a simple surrogate model for the uniaxial CP model described in Section~\ref{subsec:cp_model}, as a proof of concept for including VAE fingerprints as a parameterisation of microstructure for yielding a microstructure-dependent surrogate model.
Figure~\ref{fig:pipeline_schematic} shows a schematic of the general pipeline.

A fully connected neural network was constructed with an input layer size that matches the dimensionality of the fingerprints being operated on.
This is followed by a hidden layer of size 512, with Leaky ReLU activation, and a final 21-dimensional layer, with sigmoid activation, to output the normalised stress response for each discrete strain increment.
The Adam optimiser is used here again, operating on the mean-squared error between the prediction and true stress from the DAMASK simulations.
A learning rate of $1 \times 10^{-5}$ was used with a weight decay of $1 \times 10^{-5}$ and a batch size of 128 was used to introduce some stochasticity into the training process, rather than backpropagating after a forward pass with the entire training set (which would have fit into memory with the resources available).
This stochasticity is favourable here, as the training dataset is relatively small.


\section{Results}
\label{sec:results}

\subsection{Architecture Optimisation}
\label{subsec:architecture}

To optimise the VAE architecture, a test matrix was devised from three choices in latent space dimensionality and three choices in architecture depth.
The latent space dimensionality was tested first with a ResNet18-like architecture that was tested in previous work~\cite{White2024}.
After selection of the latent space dimensionality, different architectures were tested \,---\, namely, ResNet12, ResNet18 and ResNet34 (as defined in Section~\ref{subsubsec:3dvae}).
The learning rate and weight decay were both set to $1 \times 10^{-5}$ for the Adam optimiser, and the hidden layer before sampling $\mu$ and ln$(\sigma)$ had a size of 8192, in all cases.

\subsubsection{Latent Dimension Selection}
\label{subsubsec:dim_selection}

\begin{figure}[!ht]
    \centering
    \includegraphics[width=\textwidth]{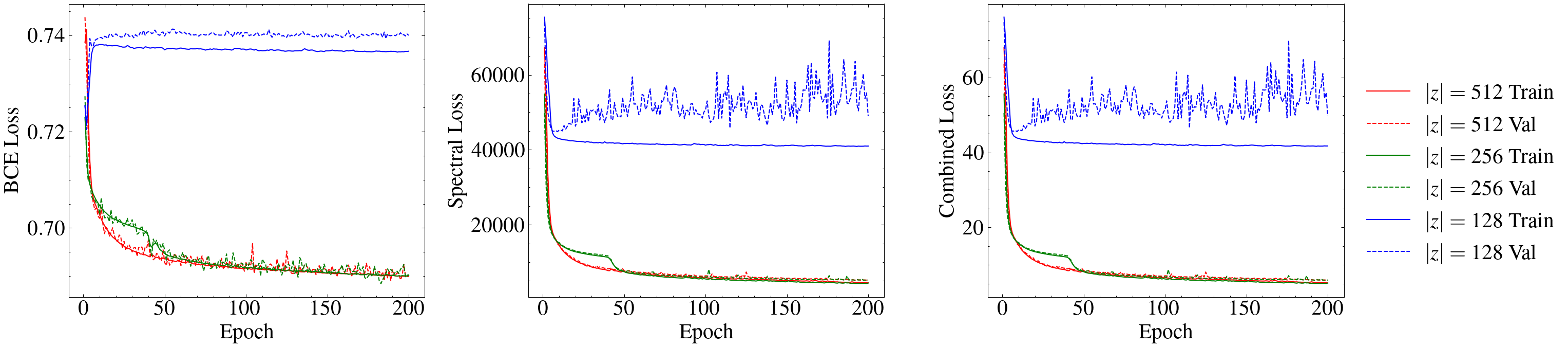}
    \caption{Loss output for optimising the dimensionality of the latent space.}
    \label{fig:latentdim_loss}
\end{figure}

The ResNet18 VAE was used to benchmark the latent space sizes as a similar architecture has been shown to perform well on 2D microstructures~\cite{White2024} and, relative to the other architectures that are being tested, ResNet18 provides a balance between model complexity and compute resource requirements.
Training was performed across two Nvidia A100 GPUs with a wall-clock time of $\sim 28$ hours to train for 200 epochs.
Figure~\ref{fig:latentdim_loss} shows the loss development throughout the training processes for 200 epochs.
The loss is decomposed into the BCE and spectral loss functions for clarity and the combined loss is also provided, which is given by Equation~(\ref{eq:loss}) with $\alpha = 0.999$.
As can be seen, this choice in $\alpha$ results in heavy weighting towards the spectral loss.

The aim here is to achieve maximum compression with minimum impact on the reconstruction accuracy.
Choosing $|z| = 256$ yields a final loss similar to that of $|z| = 512$, i.e.\ offers further compression without compromising the amount of relevant information contained in the resulting fingerprints.
Compressing further, down to $|z| = 128$, leads to significant early overfitting and the training loss also severely underperforms compared to the higher dimensional latent spaces.
Thus, a latent space with $|z| = 256$ was deemed optimal in this case and is used going forward.

\subsubsection{Network Depth Selection}
\label{subsubsec:depth_selection}

\begin{figure}[!ht]
    \centering
    \includegraphics[width=\textwidth]{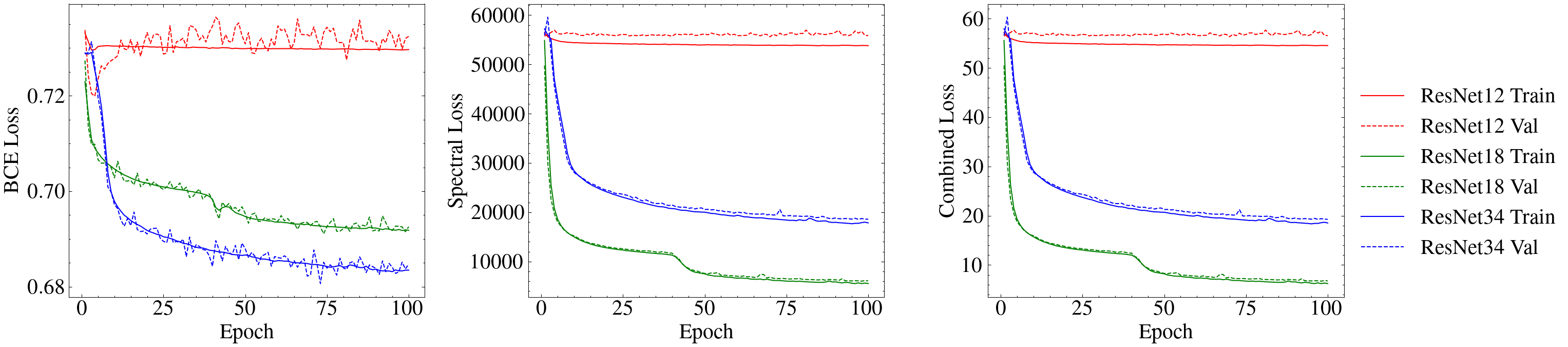}
    \caption{Loss output for optimising the ResNet architecture.}
    \label{fig:architectures_loss}
\end{figure}

Once the latent space dimensionality was fixed to $|z| = 256$, three ResNet architectures were tested.
The ResNet18 architecture that served as the baseline for selecting $|z|$ was carried forward and compared with ResNet12 and ResNet34, which are defined in Section~\ref{subsubsec:3dvae}.
The number of training epochs was reduced to 100 here, as the ResNet34 architecture is much more computationally expensive, and it was observed that the key differences in their behaviours were apparent at this point regardless.
Training was again performed across two Nvidia A100 GPUs, with wall-clock times of $\sim 9.5$, 14 and 20 hours for the ResNet 12, 18 and 34 architectures, respectively.
Figure~\ref{fig:architectures_loss} shows the loss output throughout the training process, again splitting the combined loss into its constitutive components.

Here, we see that ResNet18 offers the best performance for the combined loss.
However, as mentioned in Section~\ref{subsubsec:dim_selection}, the combined loss is heavily weighted towards the spectral component of the loss function.
Considering only the BCE loss, one could argue that ResNet34 outperforms the other architectures.
To further analyse this discrepancy, the mean misorientation between the input VEs and their corresponding reconstructions, for each architecture, was calculated after training.
The metric applied here is $\phi_4(\bm{q}_1, \bm{q}_2) \in [0, 1]$ from~\cite{huynh_metrics_2009}, given as
\begin{equation}
    \label{eq:phi4}
    \phi_4(\bm{q}_1, \bm{q}_2) = 1 - |\bm{q}_1 \cdot \bm{q}_2|,
\end{equation}
where $\bm{q}_1, \bm{q}_2$ are quaternions converted from the Euler angles in each VE.
This is calculated for each voxel in the VE.
The mean of $\phi_4$ is first calculated for each VE before the mean and standard deviation are taken across the dataset.
The results are provided in Table~\ref{tab:misorientation}.


\begin{table}[!ht]
    \centering
    \begin{tabular}{ |c|c|c| }
        \hline
        \textbf{Architecture} & \textbf{Mean misorientation (train)} & \textbf{Mean misorientation (test)} \\
        \hline
        ResNet12              & $0.070 \pm 0.006$                    & $0.070 \pm 0.006$                   \\
        \hline
        ResNet18              & $0.029 \pm 0.003$                    & $0.031 \pm 0.004$                   \\
        \hline
        ResNet34              & $0.075 \pm 0.004$                    & $0.075 \pm 0.004$                   \\
        \hline
    \end{tabular}
    \caption{Mean misorientations across training and test datasets for each network depth tested in the case where $|z| = 256$.}
    \label{tab:misorientation}
\end{table}

Despite the drop in BCE loss from ResNet18 to ResNet34, the mean misorientation is considerably lower for ResNet18.
This confirms that, although the spectral term in the combined loss function is proposed as a method for regularising the latent space~\cite{Bjork2022}, it does also act as a reconstruction loss as well.
Further analysis of the latent space, to confirm that is has been suitably regularised, is presented in Section~\ref{subsec:visualisation}.

\subsection{Reconstruction Accuracy}
\label{subsec:reconstruction}

\begin{figure}[!ht]
    \centering
    \includegraphics[width=.6\textwidth]{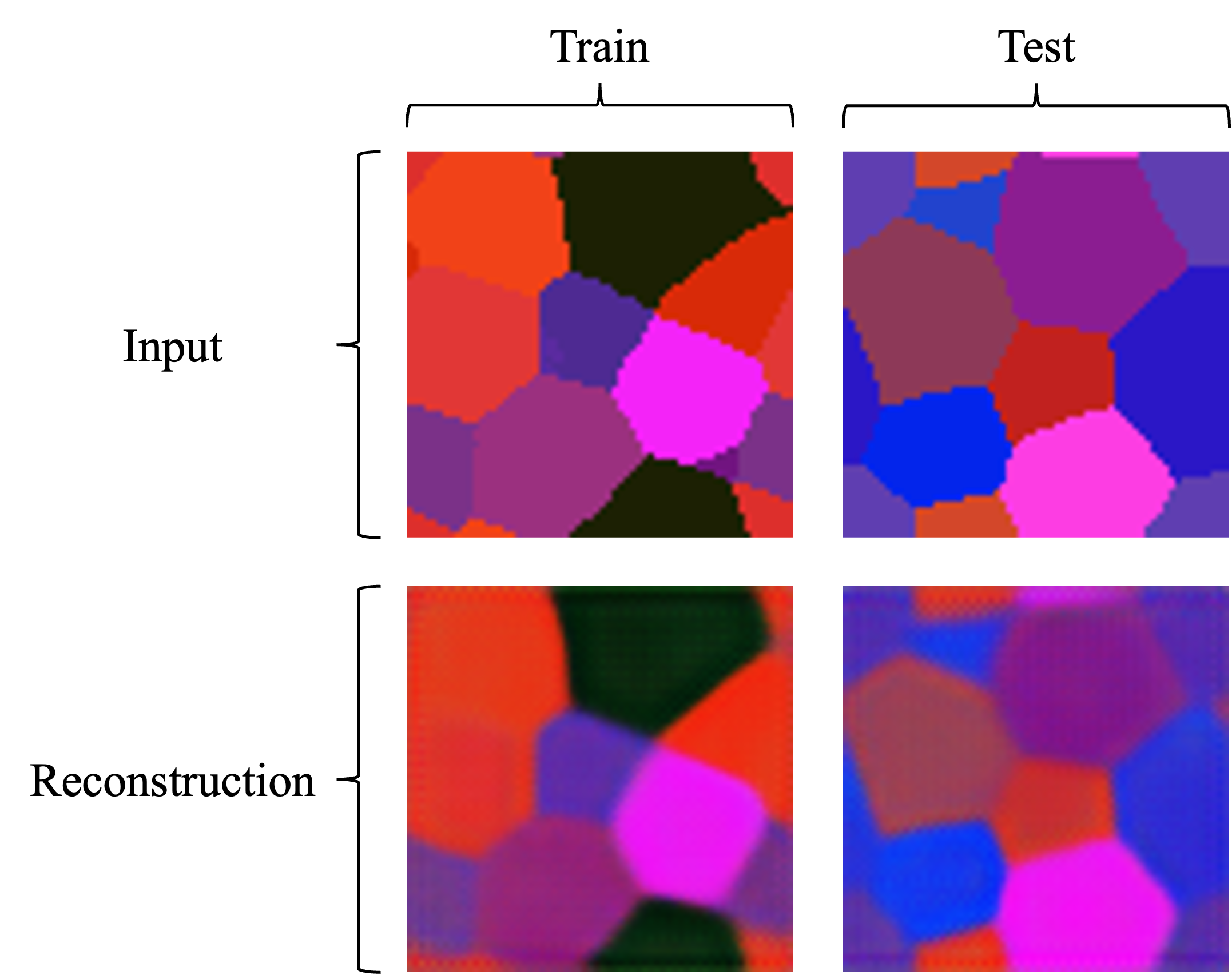}
    \caption{Representative examples of a VE slice from the training and test datasets.}
    \label{fig:reconstructions}
\end{figure}

After model selection (ResNet18 with $|z| = 256$), the trained model can output a fingerprint and a reconstruction.
The quality of the reconstruction can be used to assess the information contained in the fingerprint.
Figure~\ref{fig:reconstructions} shows a random slice from a VE from both the training and validation datasets.
As can be seen, there is some smoothing present, which is to be expected given the nature of the convolution operators that have been used to generate them.
However, the morphology and orientation information seems to be well captured in the reconstruction.
The colour map assigned to the orientations here is taken from simply mapping each normalised Euler angle to RGB values, to illustrate the data that is being seen by the VAE and to avoid any bias introduced by converting to the more typically used inverse pole figure (IPF) colouring and selecting a reference direction.
To put a metric on the reconstruction quality (that is more easily visualised than the loss function used for training), the quaternion misorientation given by Equation~\ref{eq:phi4} was used.
The outputs from this are shown in Table~\ref{tab:misorientation}, which was considered in the previous section for network depth selection.

\begin{figure}[!ht]
    \centering
    \includegraphics[width=\textwidth]{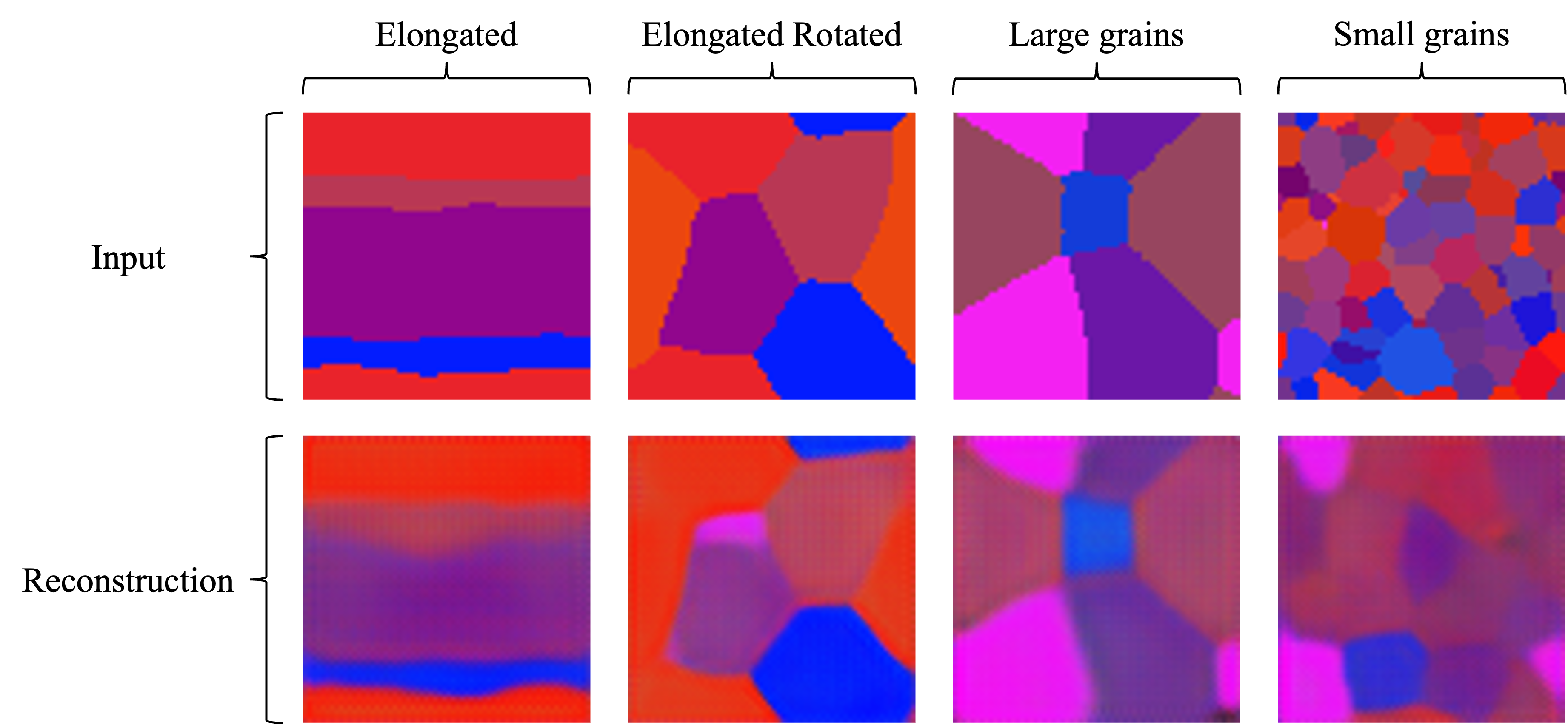}
    \caption{Representative examples of VE slices that showcase a morphological edge case, relative to the training data. Namely, large grains, small grains, elongated grains and rotated equivalent of the same elongated grains.}
    \label{fig:reconstructions_morphology}
\end{figure}

To test how well the VAE can generalise to unseen data that is not statistically equivalent to the training data, a set of edge cases were constructed, as described in Section~\ref{subsubsec:test_data}.
These focused on either morphological features, or VEs with strong textures.
In the morphological feature edge cases, grain size was included by considering relatively large and small grains compared to the grain size distribution used to sample the training data.
Elongated grains were also considered, relative to the equiaxed grains in the training data.
Figure~\ref{fig:reconstructions_morphology} shows a random sample from each of these edge cases.

\begin{figure}[!ht]
    \centering
    \begin{subfigure}{0.49\textwidth}
        \centering
        \includegraphics[width=\textwidth]{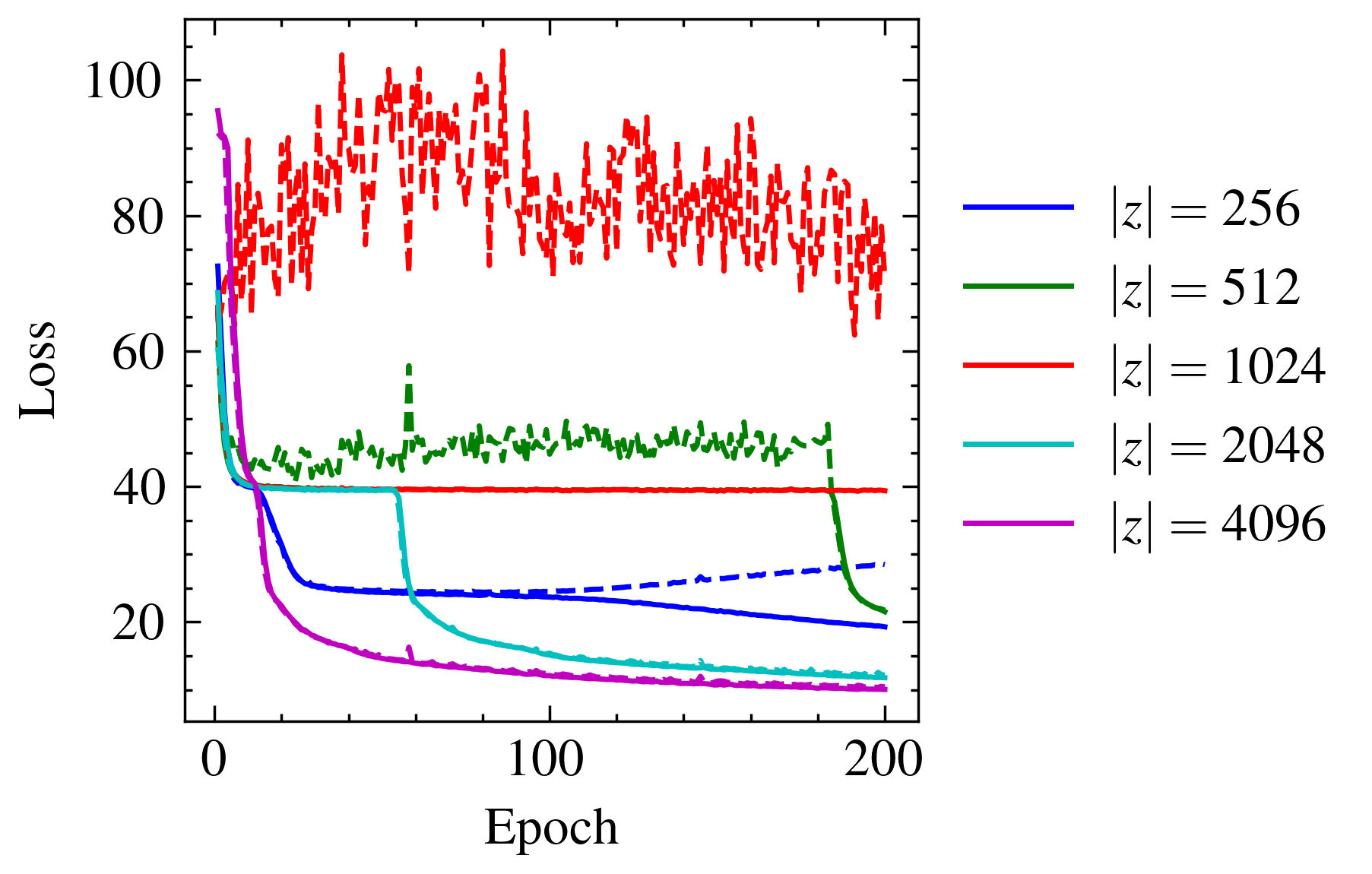}
        \caption{}
        \label{subfig:small_grain_loss}
    \end{subfigure}%
    \begin{subfigure}{0.49\textwidth}
        \centering
        \includegraphics[width=\textwidth]{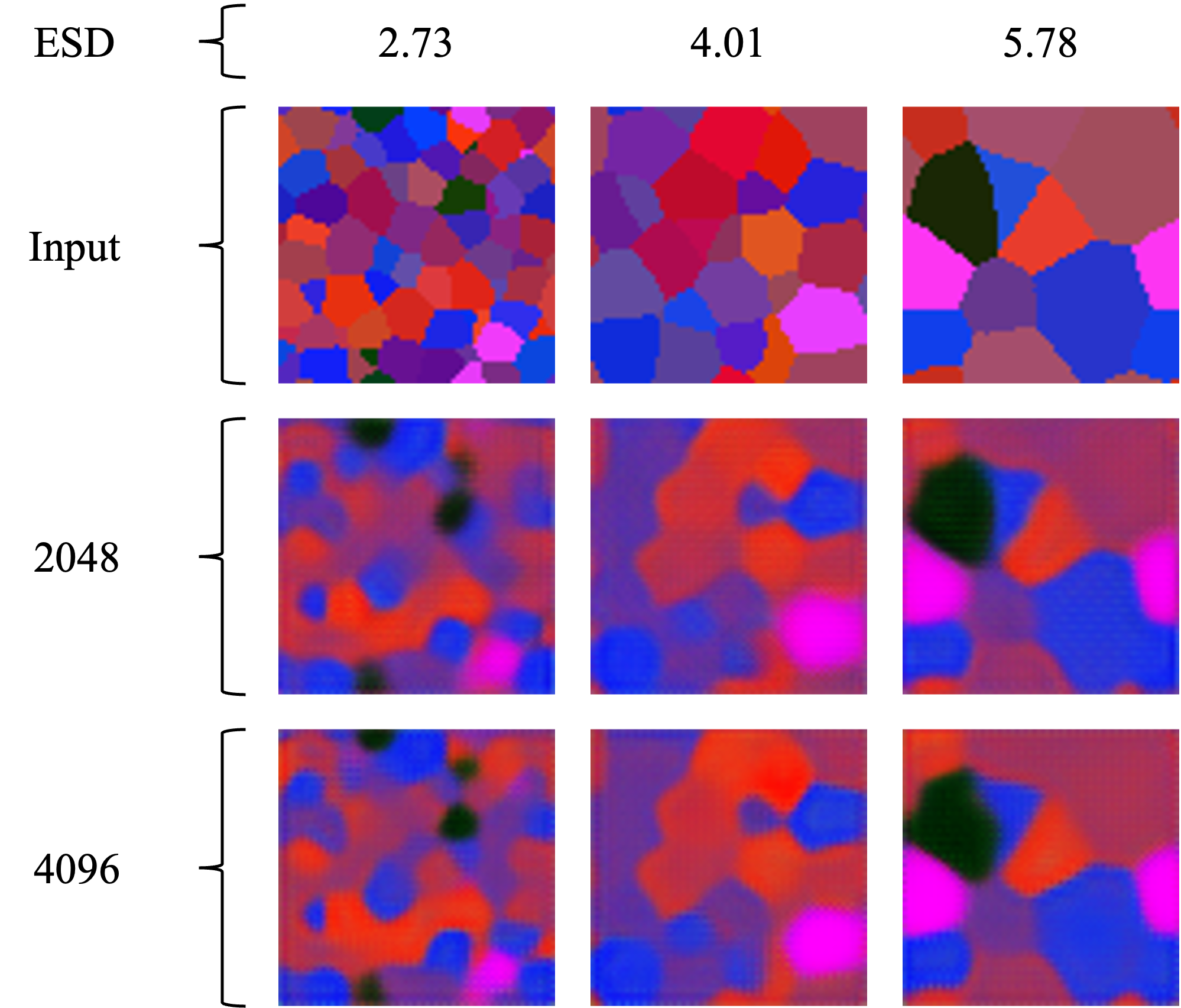}
        \caption{}
        \label{subfig:small_grain_examples}
    \end{subfigure}
    \caption{Results from retraining the VAE with a dataset containing volume elements comprising small grains comparable to the edge case described in Section~\ref{subsubsec:test_data}. (a) shows the loss behaviour throughout training for three latent space sizes and (b) shows slices from some example reconstructions.}
    \label{fig:small_grains}
\end{figure}

The large and elongated grains are reconstructed with an accuracy comparable to the training data, as shown in Table~\ref{tab:misorientation_morphology}.
However, reconstruction in the case of the small grains is clearly not representative of the input VE.
This is, in part, due to a lack of representation of small grains in the training data and, also, the fact that there is much more information contained in the VEs with smaller grains, relative to the training set, so there is less potential compression available.
If smaller grains are of interest, or there is a higher variance in grain size across the dataset, relative length scales would need to be reevaluated and the size of the latent space would need to be increased to accommodate this additional diversity in the training set.
To verify this, an additional dataset of 110,000 volume elements was generated, with an even split between volume elements with a mean ESD of 2.73 and 4.01 (smaller than the initial training dataset, which contained a mean ESD of 5.78).
100,000 were used to train the VAE from scratch, with various latent space sizes and 10,000 were reserved for validation.
The loss behaviour throughout training for each of these architectures is shown in Figure~\ref{subfig:small_grain_loss}.
In the case where $|z| = 256$, the validation loss diverges away from the training loss, showing clear signs of overfitting.
Increasing the size of the latent space starts to yield smooth behaviour at $|z| = 2048$, with no sign of overfitting and an increase in accuracy as the size of the latent space is further increased to $|z| = 4096$.
Figure~\ref{subfig:small_grain_examples} shows some example reconstructions on unseen data, which demonstrates that small grains can indeed be resolved, provided that they are suitably represented in the training data, although the smoothing effect at grain boundaries is more pronounced than with the previous dataset, due to the increased number of grain boundaries in the VE.

Generalisation up to the large grains and elongated grains from the training data is feasible due to these morphologies being obtainable by placing grains from the training data with equal orientations adjacent to each other, i.e. they can be considered as a subset of the distribution that the training data represents, whereas the small grain case is extrapolation away from the training dataset, which is notoriously challenging to deal with in machine learning models.
Suitable modification of the training data and increasing the size of the latent space can aid in encoding smaller grains, but the mean misorientation on unseen data is still higher than the cases that deal with larger grains due to the increased number of grain boundaries and the smoothing effect of the VAE on reconstructions causing blurred grain boundaries.

\begin{table}[!ht]
    \centering
    \begin{tabular}{ |c|c|c| }
        \hline
        \textbf{Morphology} & $|z|$            & \textbf{Mean misorientation} \\
        \hline
        Large grains        & 256              & $0.033 \pm 0.010$            \\
        \hline
        Small grains        & 256              & $0.063 \pm 0.001$            \\
        \hline
        Small grains        & 512              & $0.063 \pm 0.001$            \\
        \hline
        Small grains        & 1024             & $0.065 \pm 0.001$            \\
        \hline
        Small grains        & 2048             & $0.064 \pm 0.001$            \\
        \hline
        Small grains        & 256 (retrained)  & $0.061 \pm 0.001$           \\
        \hline
        Small grains        & 512 (retrained)  & $0.056 \pm 0.0008$           \\
        \hline
        Small grains        & 1024 (retrained) & $0.073 \pm 0.002$           \\
        \hline
        Small grains        & 2048 (retrained) & $0.048 \pm 0.0008$            \\
        \hline
        Small grains        & 4096 (retrained) & $0.045 \pm 0.0007$            \\
        \hline
        Elongated grains    & 256              & $0.036 \pm 0.009$            \\
        \hline
    \end{tabular}
    \caption{Mean misorientations between input VEs and their corresponding reconstructions for 100 samples from each of the morphological feature edge cases. The cases that state $|z|$ (retrained) refer to cases for which a bespoke dataset was created, with modified morphologies that are representative of the edge case, and used to train the VAE from scratch. This is to indicate that the edge cases that performed poorly (small grains) can be accounted for by modifying the training dataset.}
    \label{tab:misorientation_morphology}
\end{table}

\begin{figure}[!ht]
    \centering
    \includegraphics[width=\textwidth]{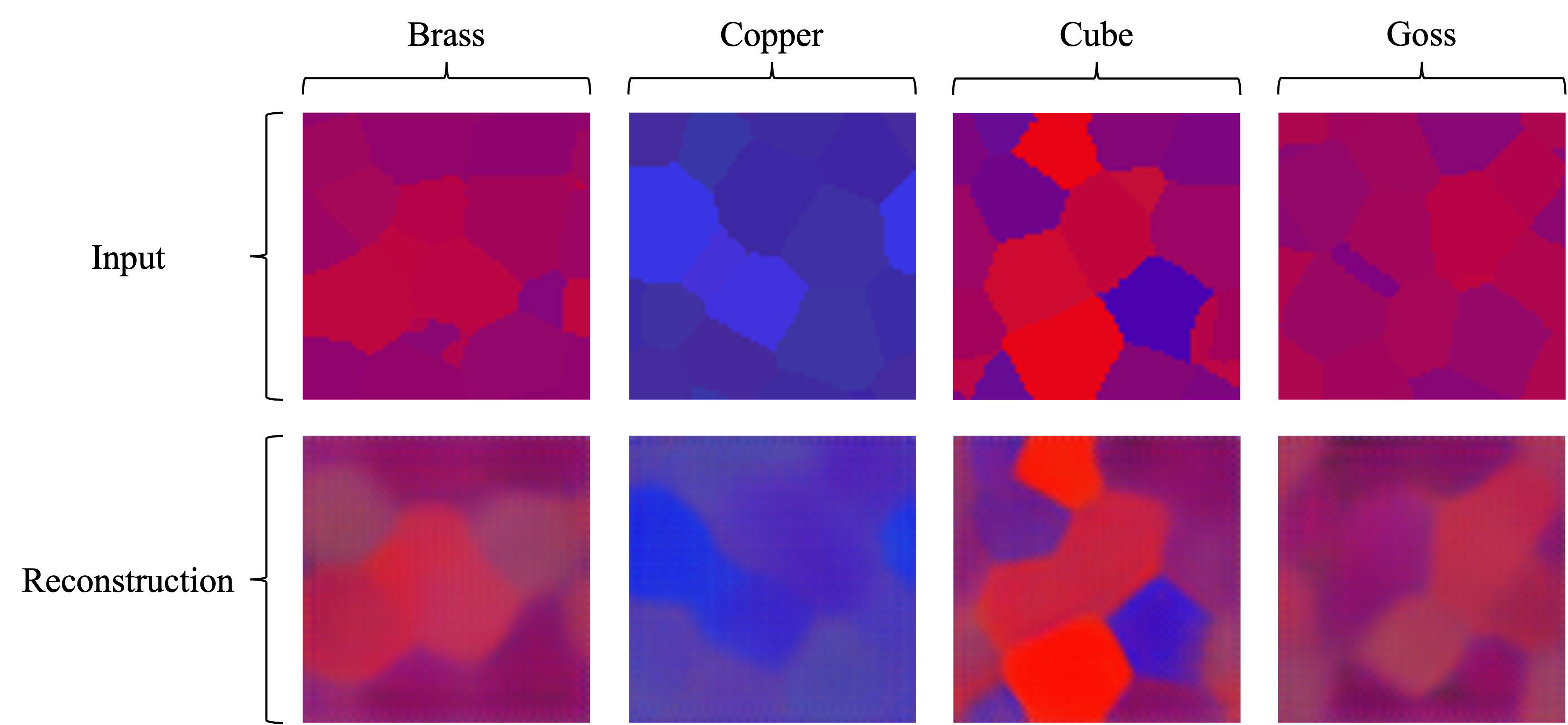}
    \caption{Representative examples of VE slices that showcase a morphological edge case, relative to the training data. Namely, brass, copper, cube and Goss.}
    \label{fig:reconstructions_texture}
\end{figure}

For the Euler angle edge cases, four common textures were considered, as described in Section~\ref{subsubsec:test_data}.
Figure~\ref{fig:reconstructions_texture} shows examples of slices from a random VE for each texture component.

As before, the mean misorientation was calculated across 100 samples for each of these edge cases and the results are provided in Table~\ref{tab:misorientation_texture}.
The reconstruction accuracy here is comparable to the training data for the brass and copper examples, with a slight drop in accuracy for the copper case.

\begin{table}[!ht]
    \centering
    \begin{tabular}{ |c|c| }
        \hline
        \textbf{Texture} & \textbf{Mean misorientation} \\
        \hline
        Brass            & $0.044 \pm 0.003$            \\
        \hline
        Copper           & $0.061 \pm 0.003$            \\
        \hline
        Cube             & $0.029 \pm 0.003$            \\
        \hline
        Goss             & $0.032 \pm 0.004$            \\
        \hline
    \end{tabular}
    \caption{Mean misorientations between input VEs and their corresponding reconstructions for 100 samples from each of the textured edge cases, with $|z| = 256$.}
    \label{tab:misorientation_texture}
\end{table}

\subsection{Latent Space Visualisation}
\label{subsec:visualisation}

\subsubsection{Linear Paths}
\label{subsubsec:paths}

\begin{figure}[!ht]
    \centering
    \includegraphics[width=\textwidth]{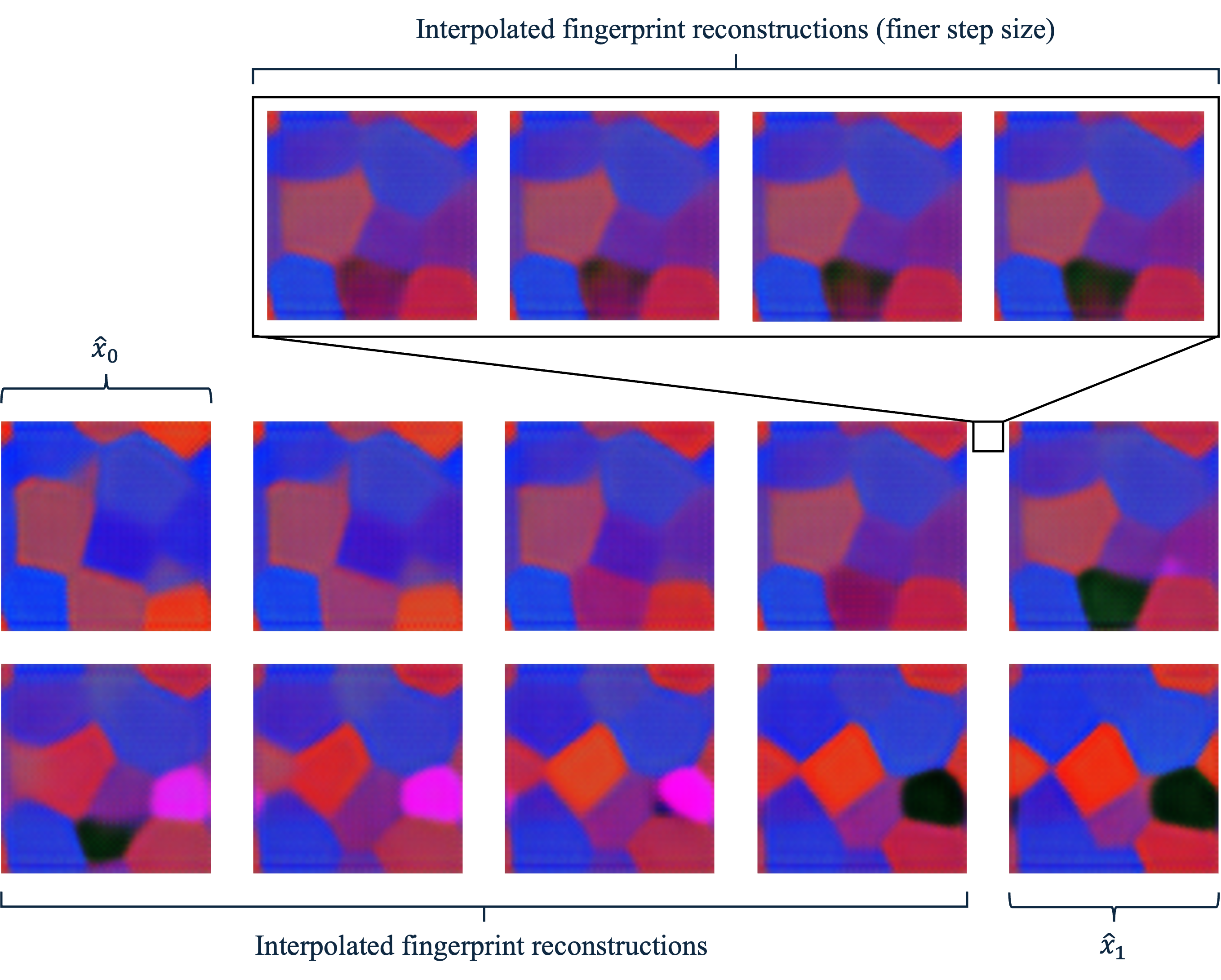}
    \caption{Reconstructions along a linear path through the latent space.}
    \label{fig:linear_path}
\end{figure}

The latent space constructed by a VAE is a continuous space, provided that it has been suitably regularised.
To visually confirm this continuity, paths through the space can be plotted.
Here, we consider a linear path with $N$ equispaced samples along the path.
Given two fingerprints, $z_1$ and $z_2$, the $N$ fingerprints ($z_n, n = 1, ..., N$) along the linear path are given by
\begin{equation}
    z_n = z_1 + \frac{n + 1}{N} z_2, \hspace{1em} n = 1, \dots, N.
\end{equation}
The $z_n$ can be passed through the trained decoder to reconstruct the microstructures, $\hat{x}$, along the path.
Figure~\ref{fig:linear_path} shows an example of such a path between two volume elements in the test dataset that were not seen during training, with $N = 8$.
The images shown are slices through the centre of each 3D reconstruction and show a smooth transition between each microstructure along the path, visually confirming the continuity of the latent space.
Continuity in the latent space is a desirable property for downstream tasks, such as parameterising effective VE properties by their fingerprint, since it guarantees smooth property mappings.
In some cases along the path shown in Figure~\ref{fig:linear_path}, there appears to be some discontinuities with individual grains suddenly switching orientation.
This is due to the step size taken between reconstructions in the latent space (smaller steps yields smoother transformations).
To verify this, we have plotted a path with a finer step size for one of these cases, which shows that the change in orientation exhibits smooth transition by engulfing the original grain, rather than there being a sudden rotation in the orientation space.

\subsubsection{Dimensionality Reduction}
\label{subsubsec:pca}

PCA~\cite{abdi_principal_2010} was used to reduce the fingerprints down to two dimensions for visualisation purposes only and not to further compress the fingerprints for downstream tasks.
The explained variance was shown to only be fully captured if the first 200 principal components were used, so this 2D representation is by no means the complete picture.
It does, however, provide some insight into which microstructural information is differentiated by the resulting fingerprints.
Spectral clustering is used to verify these findings in Section~\ref{subsubsec:clustering}.

\begin{figure}[!ht]
    \centering
    \begin{subfigure}{0.5\textwidth}
        \centering
        \includegraphics[width=\textwidth]{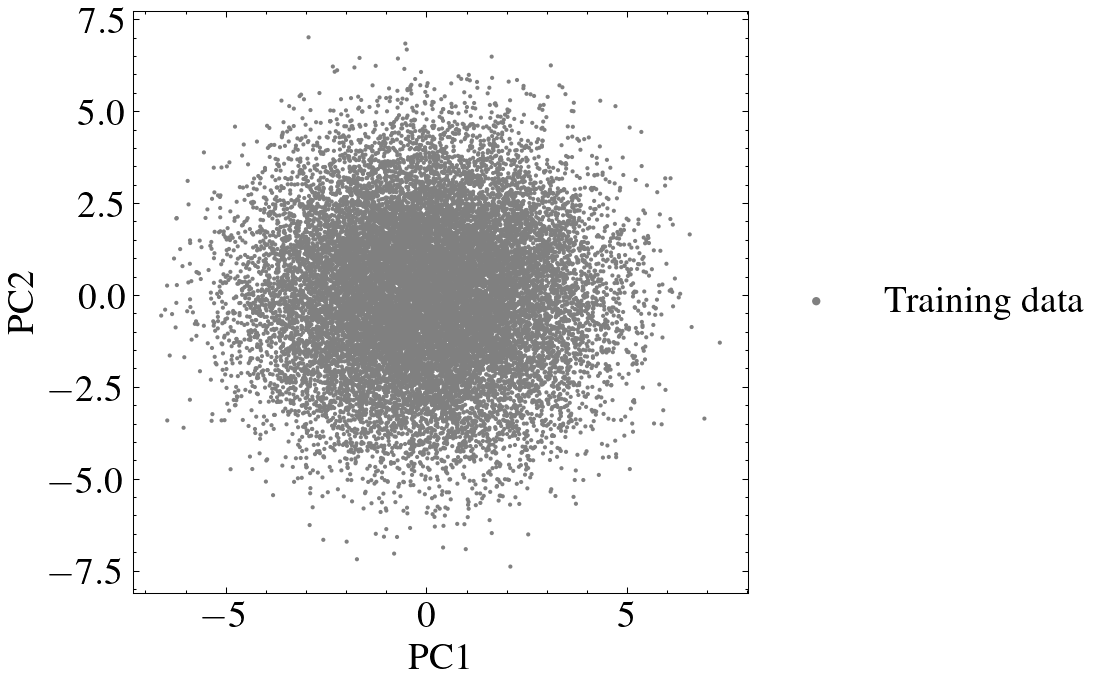}
        \caption{Training distribution}
        \label{subfig:pca_training}
    \end{subfigure}%
    \begin{subfigure}{0.5\textwidth}
        \centering
        \includegraphics[width=\textwidth]{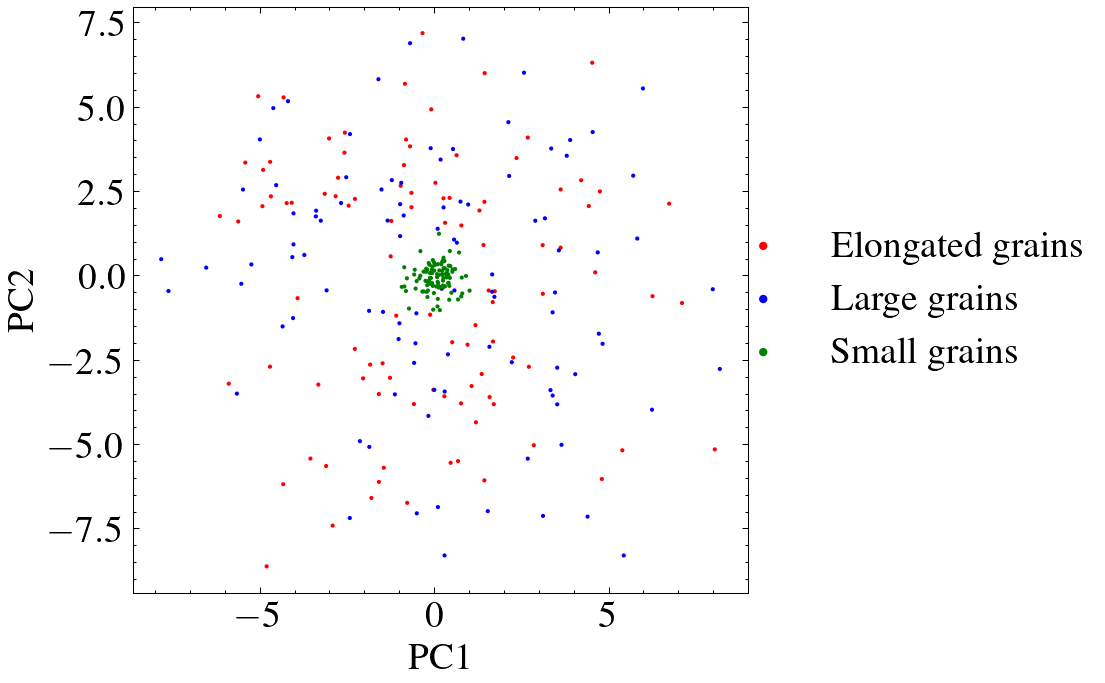}
        \caption{Morphology distribution}
        \label{subfig:pca_morphology}
    \end{subfigure}
    \centering
    \begin{subfigure}{0.5\textwidth}
        \centering
        \includegraphics[width=\textwidth]{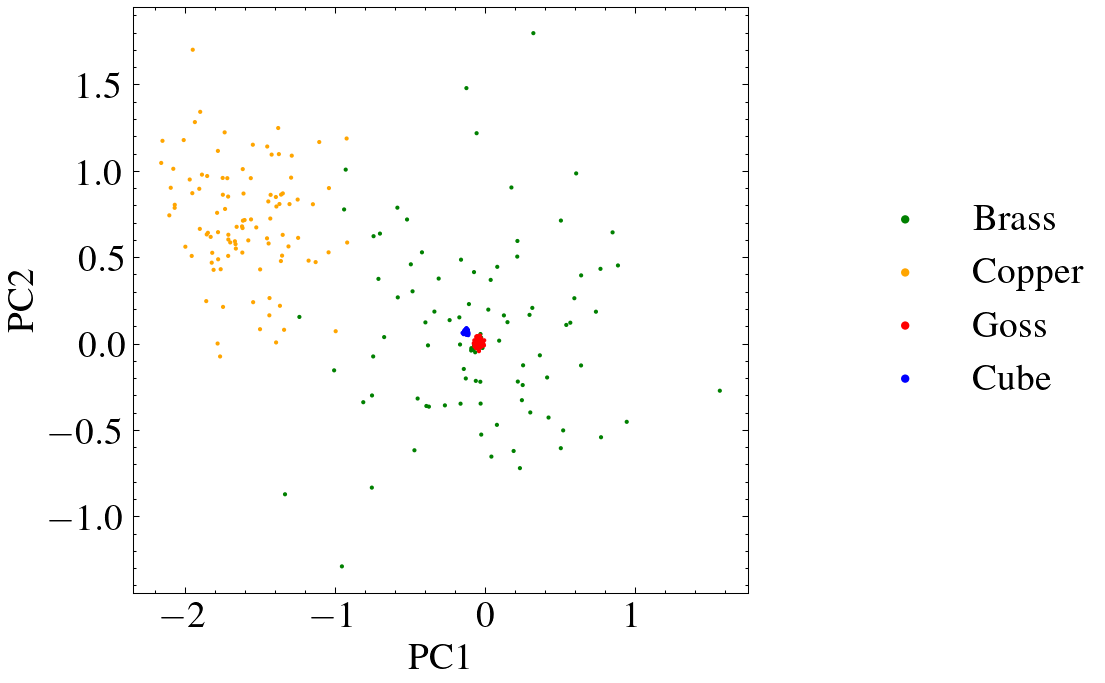}
        \caption{Texture distribution}
        \label{subfig:pca_texture}
    \end{subfigure}%
    \begin{subfigure}{0.5\textwidth}
        \centering
        \includegraphics[width=\textwidth]{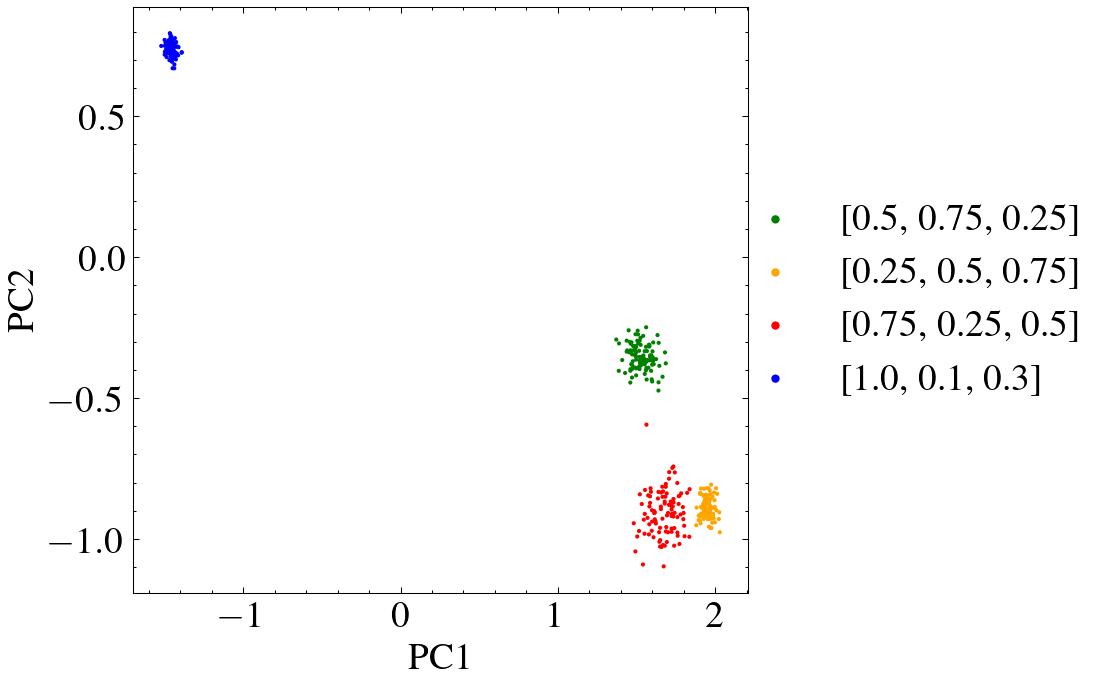}
        \caption{Orientation distribution}
        \label{subfig:pca_orientations}
    \end{subfigure}
    \caption{PCA-reduced fingerprints to highlight clustering of VEs with a strong texture and lack of clustering for morphological features.}
    \label{fig:pca}
\end{figure}

The PCA was trained on 20,000 fingerprints that were randomly sampled from the training set.
This was then used to predict the first two principal components for the edge cases.
Figure~\ref{fig:pca} shows scatter plots to illustrate the distribution of the morphological and texture edge cases across the latent space relative to the training data.
Another set of orientation-based edge cases, with a tighter variance ($0.01^\circ$), were generated to show more prominent clustering of the orientation data, and are also shown here.
These were created in much the same way as described in Section~\ref{subsubsec:test_data}, but with normalised Euler angles selected to be away from zero, rather than using common texture components.

The PCA projected training data, that was used to learn the mapping, is shown in Figure~\ref{subfig:pca_training}, to give a visual representation of the range of reasonable values for each principal component of the latent space.
The morphology edge cases (shown in Figure~\ref{subfig:pca_morphology}) appear to be randomly distributed across the latent space, except in the case of small grains.
However, the trained VAE was not able to reconstruct the small grains, either due to issues with extrapolating from the training data, or due to the minimum resolution of the VAE in general.
Morphological features being distributed in this way is contrary to previous work~\cite{White2024}, which showed that grain size can be correlated with fingerprints in the latent space.
However, greyscale images with no texture information were used in that case, with large variation in the morphologies in the training data.
This led to accurate reconstruction depending entirely on morphological information.
In our case, the morphology in the training data was entirely equiaxed, which a relatively narrow distribution of grain sizes, whereas the Euler angle space was sampled in its entirety, leading to most of the diversity in the training set arising from the textures of the VEs, rather than their morphology.

The correlation between texture and the latent space can be seen in Figures~\ref{subfig:pca_texture} and~\ref{subfig:pca_orientations}, which show fingerprints from VEs with some common textures and VEs with some specific normalised Euler angles, chosen such that they are sufficiently far away from zero.
As discussed in Section~\ref{subsubsec:test_data}, whilst generating these VEs, the orientation is forced to be strictly positive, and the absolute value is taken of any noise added to an Euler angle close to zero.
This avoids cases where the resulting angle would wrap around from 0 to $2 \pi$, but results in an artificially smaller variance in cases where the Euler angle is close to zero.
This can be seen in Figure~\ref{subfig:pca_texture}, where the Cube and Goss textures show much tighter clustering than the other cases.
The set of edge cases with normalised Euler angles further away from zero, with a relatively small variance, were also analysed in the same way to show more distinct separation and tighter clustering.
This can be seen in Figure~\ref{subfig:pca_orientations}.

\subsubsection{Spectral Clustering}
\label{subsubsec:clustering}

As shown in Section~\ref{subsubsec:pca}, the textured edge cases that contain Euler angles close to zero are very tightly clustered around the centre of the first two principal components in the latent space, such that it is difficult to visually separate them in 2D.
To confirm that the individual textures are clustering, spectral clustering~\cite{VonLuxburg2007} was applied for unsupervised classification of the textures.
Each texture was given a class label $L \in [0, 1, 2, 3]$.
Fingerprints were constructed for the textured VEs and reduced to three-dimensional vectors via the PCA model that was fit to the training data, as described in Section~\ref{subsubsec:pca}.
The clustering yielded a classification accuracy of 96.5\%, with the misclassified fingerprints corresponding to Goss misclassified as cube and vice versa.
This is due to the variance in the noise sampling to perturb the mean orientation resulting in overlaps in the sampled orientations.

The same analysis was performed with the morphology edge cases, with class labels assigned to fingerprints constructed from the small grain, large grain and elongated grain VEs.
This resulted in a classification accuracy of 100\% for the small grains, but completely random label prediction for the other two classes, due to the small grains not being meaningfully encoded and the other cases being randomly distributed across the latent space according to their orientation distribution functions, rather than their morphology.
Removing the small grain VEs and treating this as a two class clustering problem results in a classification accuracy of 53\%, confirming that there is no correlation between position in the latent space and the morphological features considered here.

\subsection{Surrogate Model Results}
\label{subsec:surrogate_results}

\begin{figure}[!ht]
    \centering
    \begin{subfigure}{0.33\textwidth}
        \centering
        \includegraphics[width=\textwidth]{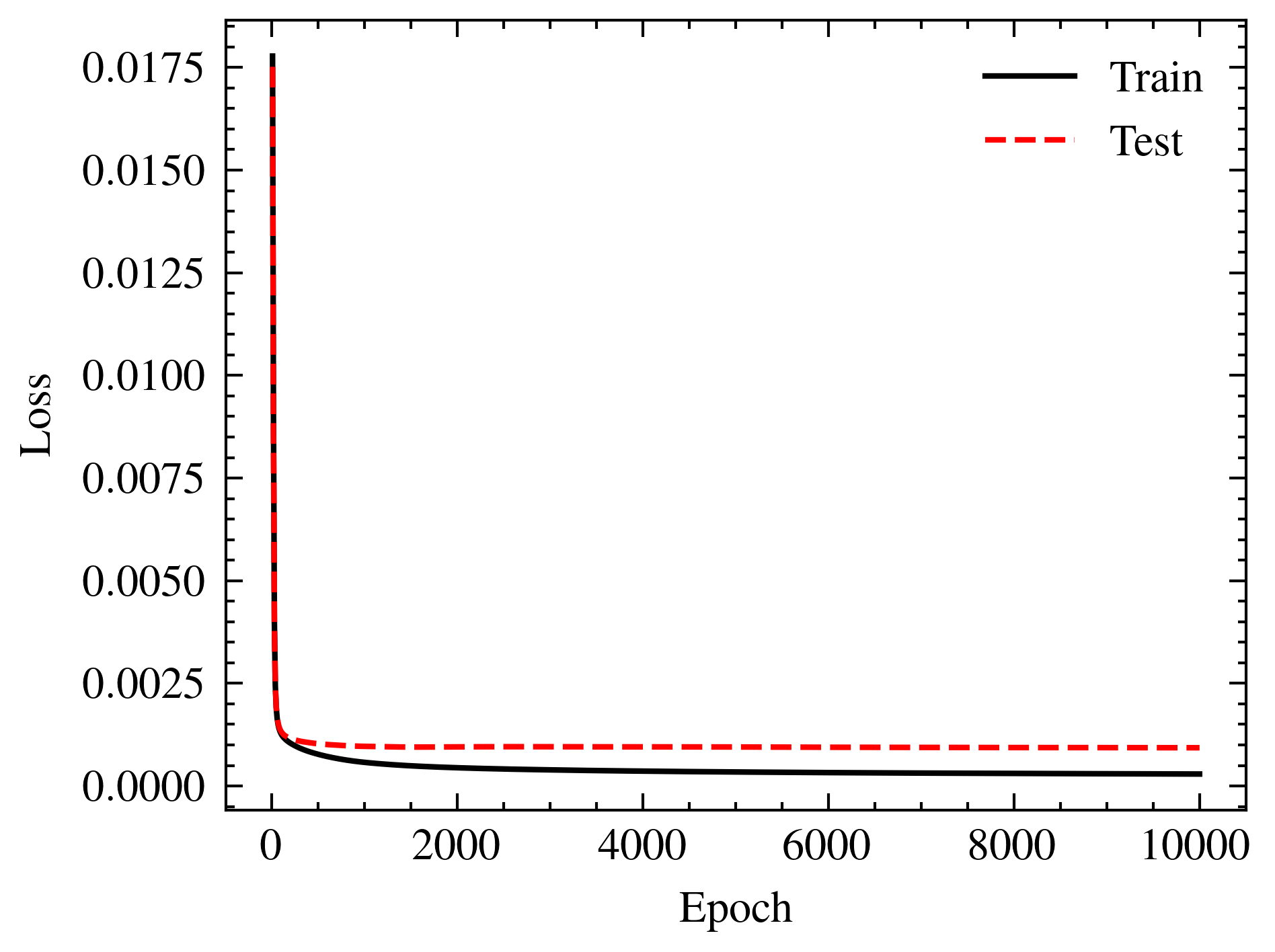}
        \caption{}
        \label{subfig:mlp_loss}
    \end{subfigure}%
    \begin{subfigure}{0.33\textwidth}
        \centering
        \includegraphics[width=\textwidth]{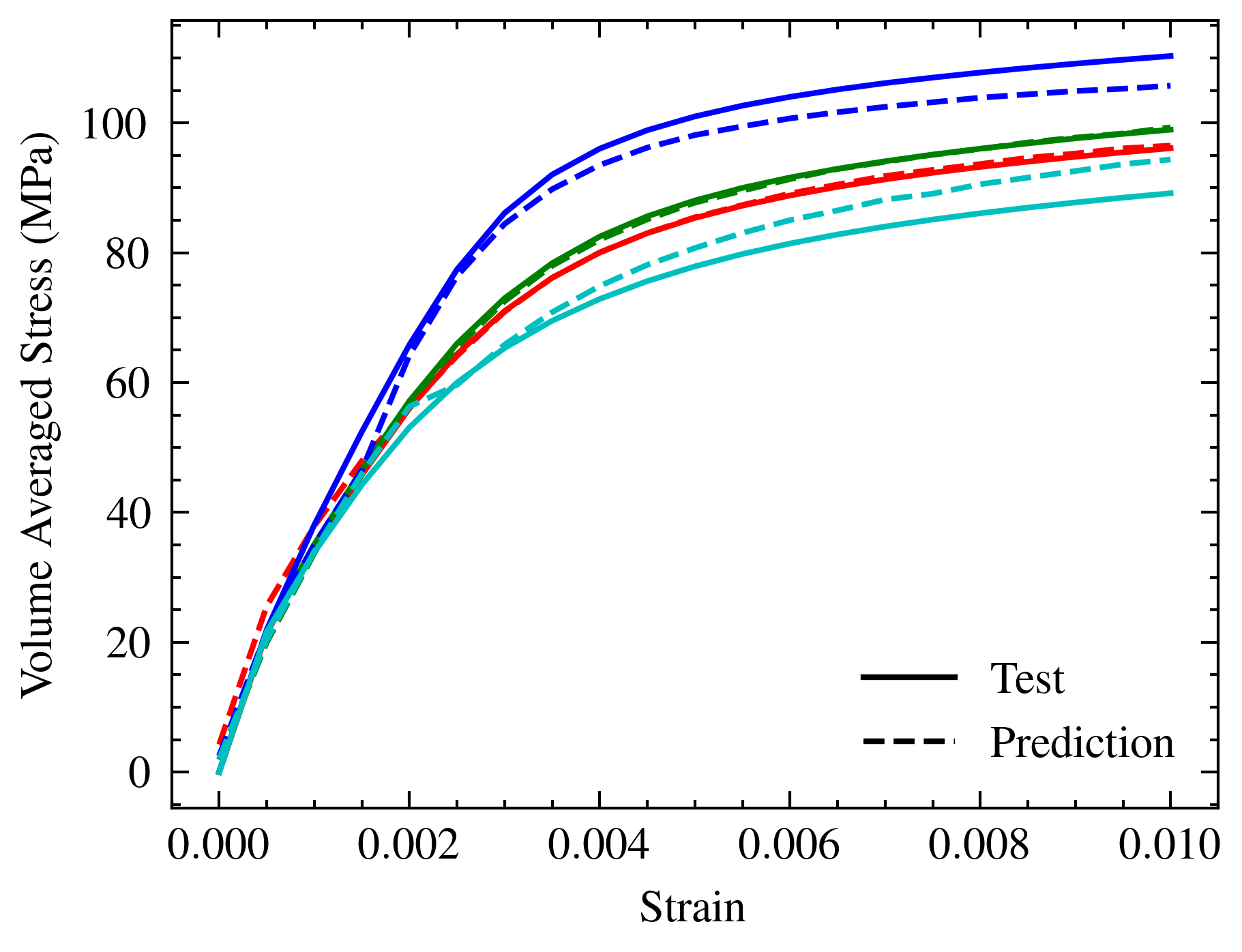}
        \caption{}
        \label{subfig:mlp_pred}
    \end{subfigure}%
    \begin{subfigure}{0.33\textwidth}
        \centering
        \includegraphics[width=\textwidth]{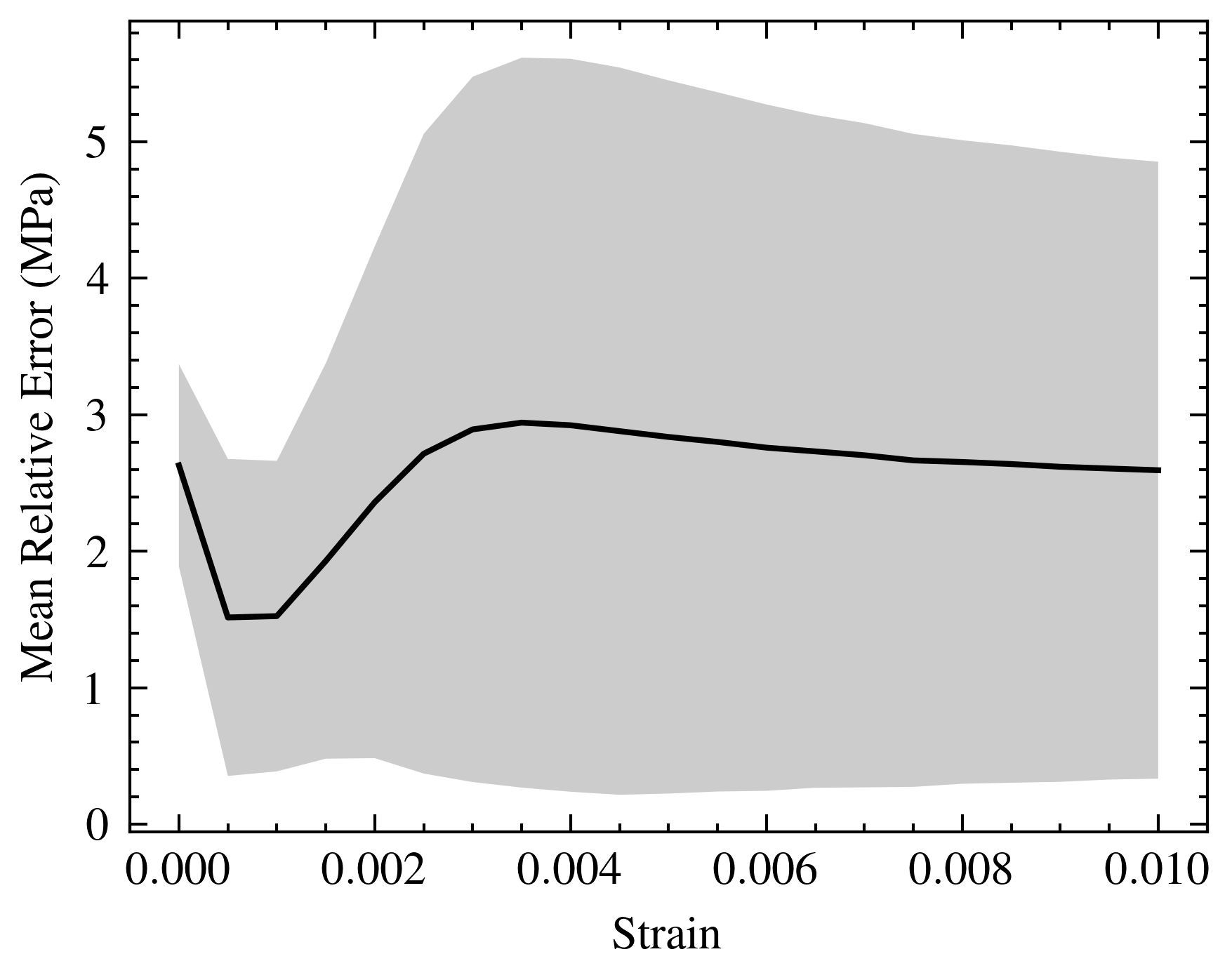}
        \caption{}
        \label{subfig:rel_err}
    \end{subfigure}
    \caption{Results from the surrogate model to show (a) the MSE loss behaviour throughout the training process and (b) some example predictions and (c) the mean relative error between the test predictions and ground truth stress with the shaded region showing the standard deviation.}
    \label{fig:mlp}
\end{figure}

The fully connected network (described in Section~\ref{subsec:surroagate}), acting as a surrogate for the uniaxial tension simulations, was trained for 10,000 epochs.
Figure~\ref{subfig:mlp_loss} provides the mean-squared error loss throughout the training process and shows smooth convergence of the prediction accuracy for both the training and test data.
The final MSE loss on the training data is $2.9 \times 10^{-4}$ and $9.3 \times 10^{-4}$ on the test data.
Once trained, the stress response predictions took $O(10^{-5})$ seconds to compute on an Apple M3 Max SoC.

Figure~\ref{subfig:mlp_pred} shows the stress response for four sample VEs picked randomly from the validation dataset.
These were selected to demonstrate that VEs with a stress response around the mean are predicted with a high accuracy, whereas stress responses nearer the extreme values suffer from worse prediction accuracy.
However, these edges are still captured somewhat, showing that there is indeed some microstructural dependence being afforded by the VAE fingerprints.
The discrepancy in accuracy for these edge cases is potentially due to a lack of representation in the training dataset.
It is also worth noting here that the stress response predictions are almost identical for all VEs for the first five strain increments.
However, these generally fall within the elastic region of the material, in which there is less of an impact from the anisotropy.
Figure~\ref{subfig:rel_err} shows the mean relative error, i.e.\ $\mathbb{E}(|y - \hat{y}|)$, where $y$ denotes the ground truth stress and $\hat{y}$ is the corresponding prediction, with the standard deviation shown as a shaded region on the plot.
The mean relative error across all strain increments is 2.75 $\pm$ 0.46 MPa, and is fairly constant across all strain increments except for a drop in error within the elastic region, where anisotropy has less of an effect.
Note that the standard deviation reported here is the standard deviation of the mean relative error across each time step and not the mean standard deviation of the relative error across each time step, which is 2.35.

\section{Discussion}

We have shown that a VAE based on ResNet18, modified with 3D convolutions, can be trained to parameterise microstructure for input into a downstream microstructure-dependent surrogate model for predicting material behaviour - in this case, a simple uniaxial CP simulation with fixed loading.
It is foreseen that the pipeline presented here could be adapted by switching out the predictive surrogate with any model of interest, without the need to retrain a microstructure encoder, as one would need to do for a full end-to-end model.
The VAE was shown to accurately reconstruct microstructure VEs with an Euler angle triplet at each voxel, however, a relatively large dataset was required for VAE training and care must be taken to account for the symmetries involved.
The reconstructions are used to train the encoder in a self-supervised manner, which can be used to parameterise microstructure in a predictive model, and the trained decoder is not used in any downstream tasks, other than to provide some visual aid towards understanding the latent space that has been constructed.
To ensure that the training process is adaptable to all crystal types, symmetry was removed as a preprocessing step.
It is possible to construct a loss function that takes the symmetries into account, but the Euler angle fundamental zones have a non-trivial geometry, so it is usual to project the Euler angles to quaternions before mapping to the fundamental zone, and this is an expensive operation to perform repeatedly throughout the training process.
Furthermore, dealing with symmetries in the loss function can result in a discontinuous loss, which should be avoided for backpropagation.
Dealing with symmetries as a preprocessing step allows for the loss to be continuous, but it is important to note that a VAE trained on a dataset with a specific fundamental zone, say cubic or hexagonal, is restricted to operating on unseen data constructed from the same fundamental zone.
If symmetry is not considered at all, the VAE can be trained to operate on any crystal structures, making it more generalisable.
However, the latent space then views symmetric variants of an Euler angle as completely different angles and, as such, textured VEs may not cluster in the latent space as one might expect.
Additionally, this would result in learning a significantly more complex mapping, requiring greater optimisation effort as well as larger model parameterisation.
Euler angles were used primarily as they allow for standard convolutions to be applied to each angle in the triplet independently, as separate image channels.
They are also simple to normalise and any reconstructions output from the VAE would be valid and applicable to being input into the loss function.
If, say, quaternions were used instead, the reconstructions would not necessarily yield valid quaternions without some additional constraint in the loss function to force the $L_2$ norm to equal one.

Some key parameters related to the VAE architecture (latent space dimensionality and ResNet depth) were optimised with a small grid search, resulting in a dimensionality reduction of over $3000 \times$ and a mean misorientation between inputs and reconstructions of $0.031 \pm 0.004$ for unseen VEs that are statistically equivalent to the training data.
This grid search was performed over more thorough optimisation algorithms due to the computational cost of training the VAE with the resources available.
Once trained, the VAE was tested on some edge cases.
These included grain morphologies and textured VEs that were not included in the training dataset.
For the morphology cases, it was shown that the VAE generalises well to VEs with larger grains than those sampled to construct the training data and, also, to elongated grains as opposed to equiaxed grains, with mean misorientations of $0.033 \pm 0.010$ and $0.036 \pm 0.009$, respectively, which is comparable to the reconstruction accuracy on the training data.
However, in cases where the grain size was much smaller than in the training dataset, the reconstructions were very poor, with a mean misorientation of $0.063 \pm 0.001$.
This is due to the larger and elongated grains effectively being a subset of the training data, which can be constructed by sampling grains that are statistically equivalent the training data, but with adjacent grains sharing the same crystallographic orientation.
Whereas the smaller grains are only achievable through extrapolation away from the training dataset.
Therefore, the VAE generalises well to unseen within-distribution samples, but not outside the distribution that it is trained on.
The VAE was retrained on a dataset containing smaller grains with two distinct mean ESDs to add more variability to the dataset and this showed that suitable representation in the training data and increasing the size of the latent space can lead to better encoding and reconstruction of the small grain edge cases.

When keeping the grain morphology consistent with the training data and considering textured VEs as edge cases, the VAE generalises well in all cases with mean misorientations comparable to those for the training data.
Despite these textures not explicitly existing in the training data, the orientations sampled to construct them do exist in the training data, and so this is more reliant on interpolation of the latent space, which works well.
Therefore, to construct a more general latent space for microstructure, sampling orientations uniformly across the Euler space leads to good generalisation, but higher variance on the grain size distribution across the training data is advisable, although this will likely result in the requirement for a larger training dataset.

To visualise the distribution of microstructural features across the latent space, we utilised dimensionality reduction, via PCA, down to two dimensions, to get a birds eye view over the space.
This is by no means a complete representation of the space, with the explained variance only being fully captured by considering the first 200 dimensions (which suggests that we are close to achieving maximum compression with the VAE constructing a latent space containing 256 dimensions).
However, it does show strong clustering of textured VEs, which is to be expected given the preprocessing to map the Euler angle triplets into a fundamental zone, although the morphological features appear to be randomly distributed across the latent space \,---\, likely due to the lack of variability in grain morphology in the training data.
This was confirmed with spectral clustering, which provides near perfect classification accuracy for textured VEs, compared to effectively random predictions for the morphological features.
We also briefly looked at a linear path through the latent space for visual confirmation that the latent space is smooth and continuous \,---\, a desirable trait for a generalisable VAE.

Once hyperparameters have been selected, the heavyweight training of the VAE is only required to be performed once and allows for a more lightweight surrogate model to be appended, which can be swapped out depending on the task.
This is much more efficient than training a new end-to-end model each time a new predictive model is of interest.
In this case, the trained VAE was used in conjunction with DAMASK to construct a new training dataset that consisted of a set of fingerprints with associated volume-averaged stress responses under uniaxial tension, for a set of magnesium VEs, as a proof of concept for applying VAE fingerprints to produce a microstructure-dependent surrogate model.
A fully connected network was trained as a microstructure-dependent surrogate for the CP simulations.
This was shown to be somewhat effective, with the spread in stress response across the dataset being captured, although, the model would benefit from a larger training dataset and/or a more carefully curated dataset to yield a more uniform sampling from the stress space.
Active learning could be employed to this effect, but it is difficult to identify microstructures that would result in edge cases a priori.
However, given a large enough dataset, it could be sampled to remove bias towards the mean stress response.
Once trained, the stress response can be predicted in $O(10^{-5})$ seconds, yielding a speed-up of $O(10^{8})$ over the CP simulation, with a mean relative error of 2.75 $\pm$ 0.46 MPa.
This speed-up would become even more prominent if finer stress increments were used, although the course strain increments used here were deemed favourable in this case to enable a relatively large dataset to be constructed for training.
It is important to note that, although we have established microstructure dependence here, we have neglected any dependence on the load path and developing a surrogate for predicting stress response from a given load path remains an avenue for further work.
However, the advantage of training a VAE separately to the predictive model, rather than a fully end-to-end model, is that the trained VAE encoder is application agnostic and can be bolted on to any feasible microstructure-dependent surrogate model as a preprocessing step.
This could be achieved by training an RNN that takes a VAE fingerprint as input, to encode the initial microstructure configuration into the hidden state of the RNN, alongside inputting the strain tensor at each recurrent block.
This could be trained to predict the stress response for a given microstructure VE and prescribed strain path.
The speed-up that this would afford, compared to running a full CP simulation, would enable microstructure-dependent stress responses to be embedded into larger scale finite element models, at each node individually, bridging the gap between crystal plasticity and the component scale models.

\section{Conclusions}
\label{sec:conclusions}

\begin{itemize}
    \item VAEs can be trained on 3D microstructure VEs and yield accurate reconstructions with over $3000 \times$ dimensionality reduction from $64^3$ to 256.
    \item The trained VAE can be generalised to successfully encode larger grains than those in the training set and elongated grains, despite only being trained on equiaxed grains. This does not hold true when considering grains that are smaller than those in the training set.
    \item There is strong clustering of textured VEs in the latent space, but morphological features appear to be randomly distributed across the space.
    \item The trained VAE can be used as a preprocessing step for parameterising microstructure for microstructure-dependent surrogate models.
    \item A fully connected network was successfully trained on the VAE fingerprints to produce a microstructure-dependent surrogate model for uniaxial tension simulations. However, the prediction accuracy for the edge cases is lower than those around the mean stress response. This is a data-intensive model that either requires a larger dataset, or more careful sampling of the space.
\end{itemize}

\section*{Data Availability Statement}

The VAE and surrogate model are available on GitHub~\cite{mike_white_lightning-vae3d_nodate}.
The datasets and model checkpoints used to generate the results presented can be found on Zenodo~\cite{michael_white_3d_nodate}.

\section*{Acknowledgements}

This work has been funded by STEP, a major technology and infrastructure programme led by UK Industrial Fusion Solutions Ltd (UKIFS), which aims to deliver the UK’s prototype fusion power plant and a path to the commercial viability of fusion.

\bibliographystyle{elsarticle-num}
\bibliography{references}

\end{document}